\documentclass[sigconf]{acmart}
\setcopyright{none}
\renewcommand\footnotetextcopyrightpermission[1]{} 
\usepackage{multirow}
\usepackage{subcaption}
\usepackage{algorithm,algorithmic}
\usepackage{xcolor,colortbl}
\usepackage[export]{adjustbox}

\AtBeginDocument{%
  }

\newtheorem{remark}{Remark}
\newtheorem{definition}{Definition}
\newtheorem{theorem}{Theorem}

\newtheorem{proposition}{Proposition}
\newtheorem{assumption}{Assumption}

\newcommand{\argmax}{\arg\!\max}

\newcommand{\maximize}[1]{\underset{{#1}}{\text{maximize}}}

\usepackage{alphalph,etoolbox} 
\patchcmd{\subequations}{\alph{equation}}{\alphalph{\value{equation}}}{}{}

\newcommand{\yw}[1]{{\color{black}#1}}
\newcommand{\yq}[1]{{\color{black}#1}}

\begin{document}

\title{Online Energy Storage Arbitrage under Imperfect Predictions: \\ A Conformal Risk-Aware Approach} 

\author{Yiqian~Wu}
\affiliation{%
  \institution{Columbia University}
    \country{United States of America}
}
\email{yiqian.wu2@columbia.edu}

\author{Ming~Yi}
\affiliation{%
  \institution{Columbia University}
    \country{United States of America}
}
\email{my2826@columbia.edu}

\author{Bolun~Xu}
\affiliation{%
 \institution{Columbia University}
  \country{United States of America}
}
 \email{bx2177@columbia.edu}

\author{James~Anderson}
\affiliation{%
  \institution{Columbia University}
   \country{United States of America}
}
\email{james.anderson@columbia.edu}

\begin{abstract}
\yw{This work proposes a conformal approach for energy storage arbitrage to control the downside risk arising from imperfect price forecasts.
Energy storage arbitrage relies solely on predictions of future market prices, while inaccurate price predictions may lead to significant profit losses.} Based on conformal decision theory, we describe a controller that dynamically adjusts decision conservativeness through prediction sets without distributional assumptions. To enable online calibration when online profit loss feedback is unobservable, we establish that a temporal difference error serves as a measurable proxy. Building on this insight, we develop two online calibration strategies: prediction error-based adaptation targeting forecast accuracy, and value error-based calibration focusing on decision quality.  Analysis of the conformal controller proves bounded long-term risk with convergence guarantees in temporal difference error, which further effectively manages risk exposure in potential profit losses. Case studies demonstrate superior performance in balancing risk and opportunity compared to benchmarks under varying forecast conditions. 
\end{abstract}


\maketitle
\fancyhead{}

\section{Introduction}\label{Sec:Intro}
\yw{Energy price arbitrage has become the main service for grid-scale batteries}, with more than 15 GW deployed in California~\cite{CaliforniaStorage2024} and 6~GW in Texas~\cite{TexasStorage2025}. Price arbitrage of energy storage fundamentally relies on accurate predictions of future market conditions and opportunity values. 
However, \yw{due to the intermittency of renewable energy, demand fluctuations, network congestion, and complex market dynamics, electricity markets exhibit high volatility, making accurate forecasting extremely challenging.
Prediction errors in arbitrage decisions not only reduce immediate profit but also propagate} through sequential operations, as current charging or discharging choices affect the future state of charge (SoC) and available actions. The decision value is only revealed retrospectively after the full arbitrage window, \yq{for instance, when evaluating annual profit once all prices within the year are realized}, making real-time algorithm calibration particularly challenging. These issues significantly impact the economic viability of energy storage \yw{and underscore the need for uncertainty-aware, sequential decision-making frameworks}.

Existing approaches to managing risk and prediction uncertainty in storage arbitrage remain inadequate. \yw{Stochastic and robust optimization methods often rely on restrictive distributional assumptions or lead to overly conservative strategies that forgo profitable opportunities~\cite{wu2025}. Online learning approaches require real-time feedback, which is unavailable in arbitrage settings. In contrast, conformal prediction and conformal decision theory provide distribution-free uncertainty quantification with formal coverage guarantees~\cite{gibbs2021adaptive,angelopoulos2022,angelopoulos2023risk,feldman2023achieving,angelopoulos2023,lekeufack2024,li2025neural}, enabling decision-makers to explicitly account for prediction uncertainty without assuming specific probability distributions. While recent studies have explored conformal calibration and risk-aware training frameworks~\cite{yeh2024,yeh2025}, they focus on static, day-ahead settings~\cite{donti2017} and overlook the temporal dependencies and evolving uncertainty of sequential arbitrage. Further research is needed to extend conformal methods for real-time, adaptive energy storage arbitrage under uncertainty.}

\yw{This paper addresses the critical challenge of managing imperfect predictions in online energy storage arbitrage when profit loss feedback is unobservable by developing a novel conformal risk-aware decision-making approach.}
This work makes the following  contributions:
\begin{itemize}
    \item Development of a  conformal risk-aware pipeline that dynamically calibrates online arbitrage decision-making to manage risk exposure from imperfect predictions, providing statistical guarantees without distributional assumptions.
    \item Demonstration that the temporal difference error can be used as a measurable proxy for unobservable value error, thus enabling automatic algorithm adaptation leveraging available online information while ensuring the cumulative temporal risk is within predefined bounds.
    \item Design and analysis of two distinct decision-making conservativeness calibration strategies including
    \yw{a heuristic method based on temporal prediction error and a decision-focused method based on temporal value error.}
    \item Demonstration through case studies that the proposed approach effectively balances risk and opportunity under varying levels of forecast accuracy, recovering up to 47\% of optimal profit under poor predictions while remaining competitive in cases with accurate forecasts, with flexibility across varying risk preferences.
\end{itemize}

The remainder of this paper is organized as follows: Section~\ref{sec:related} reviews related work. Section~\ref{sec:Prob} formulates the storage arbitrage problem, formalizes the challenges caused by price uncertainty and outlines the proposed approach. Section~\ref{sec:Meth} introduces the conformal risk-aware arbitrage framework and develops the core theoretical results. Section~\ref{sec:loss} presents the online calibration strategies. Section~\ref{sec:Case} provides comprehensive case studies. 

\section{Related Work}\label{sec:related}
\subsection{Risk-Aware Energy Storage Arbitrage}

Traditional energy storage arbitrage strategies typically \yw{adopt a risk-neutral objective, aiming to maximize expected profit using approaches such as stochastic optimization~\cite{krishnamurthy2017energy}, stochastic dynamic programming~\cite{zheng2022, megel2015stochastic}, reinforcement learning~\cite{wang2018energy}, and hybrid methods that combine machine learning with optimization~\cite{baker2023transferable, wu2024, yi2025, yi2025decision}. These risk-neutral strategies can effectively capture theoretical arbitrage opportunities and achieve near-optimal performance under accurate forecasts. Yet, electricity prices, which do not follow any analytical stochastic process, are inherently volatile due to network congestion and the need to maintain a balance between instantaneous supply and demand. Consequently, electricity price forecasts are often inaccurate. As a result, energy storage arbitrage participants are  exposed to downside risk arising from imperfect predictions, and incorporating risk considerations into arbitrage optimization is critical.}

Recent work in storage arbitrage under uncertainty has explored various risk metrics. These approaches  either rely on assumptions about probability distributions of market uncertainties or produce overly conservative strategies. For example, traditional risk management approaches, such as Conditional Value-at-Risk (CVaR)~\cite{jabr2005robust}, rely heavily on scenario generation techniques, but are inconvenient for practical implementations given the high volatility and the non-stationary nature of electricity prices.
Stochastic optimization approaches, such as robust and chance-constrained optimization methods, while providing worst-case guarantees, tend to be overly conservative and may miss profitable opportunities~\cite{wu2025}. 
Online uncertainty management approaches, such as online convex optimization~\cite{hazan2022introduction, chen2015}, require observable ground truth or performance feedback, but in energy storage arbitrage, the true value of decisions only becomes apparent retrospectively after the entire arbitrage window, making real-time algorithm adaptation extremely challenging. Moreover, the performance of such algorithms can be further compromised due to storage operational constraints. 
\yw{A summary of these representative algorithms is provided in Table~\ref{tab:algorithms}.}

\begin{table}[!t]
    \centering
    \captionsetup{justification=centering, labelsep=period, textfont=sc}
    \caption{\vspace{0mm}Comparison of baseline algorithms for risk-neutral and risk-aware storage arbitrage with two variants of the proposed approach  \vspace{0mm}}
  {\footnotesize
    \begin{tabular}{c|c|c|c|c}
        \toprule
 Strategy                           & Risk-Aware   & \begin{tabular}[c]{@{}c@{}}Distribution\\ Free \end{tabular}  & \begin{tabular}[c]{@{}c@{}}Online \\ Calibration\end{tabular} & Reference\\ \midrule
Risk Neutral             & $\times$                     & $\checkmark$            &  $\times$      & \cite{zheng2022,baker2023transferable} \\ \midrule
CVaR           & $\checkmark$  &   $\times$         &         $\times$    & \cite{jabr2005robust}      \\\midrule
Chance Constrained  & $\checkmark$&    $\times$     &   $\times$       & \cite{wu2025}            \\\midrule
Robust Optimization         & $\checkmark$ &  $\checkmark$  &     $\times$         & \cite{wu2025}            \\ \midrule
Switching Cost   & $\checkmark$                      & $\checkmark$        &     $\times$  & \cite{chen2015}     \\ \midrule
\cellcolor{gray!25}\textbf{CC--prediction err} & \cellcolor{gray!25}$\checkmark$       &    \cellcolor{gray!25}$\checkmark$        & \cellcolor{gray!25}$\checkmark$     & \cellcolor{gray!25}this work        \\ \midrule
\cellcolor{gray!25}\textbf{CC--value err}      & \cellcolor{gray!25}$\checkmark$            &     \cellcolor{gray!25}$\checkmark$    & \cellcolor{gray!25}$\checkmark$     & \cellcolor{gray!25}this work        \\  \bottomrule
\end{tabular}
}
    \label{tab:algorithms}
\end{table}

\subsection{Conformal Prediction and Decision Theory}
Conformal prediction provides distribution-free uncertainty quantification by constructing prediction sets (a set of possible predictions) with finite-sample coverage guarantees~\cite{papadopoulos2002,vovk2005,lei2013,angelopoulos2022}. \yw{A brief summary of conformal prediction is provided in Appendix~\ref{app:conformal_prediction} for reference.} Unlike traditional statistical methods that rely on distributional assumptions, conformal prediction only requires the exchangeability of data points, making it broadly applicable across different domains.
Conformal risk control~\cite{angelopoulos2023risk} extends conformal prediction to control arbitrary bounded loss functions beyond simple miscoverage. This framework enables risk control for complex decision-making scenarios where traditional accuracy metrics may not capture the full cost of incorrect decisions.

Another line of work~\cite{gibbs2021adaptive,feldman2023achieving,angelopoulos2023,lekeufack2024} circumvents the assumptions on data exchangeability by adopting a control perspective that dynamically adapts prediction sets to real-time errors to achieve desired target bounds, extending the framework to online settings~\cite{li2025neural}.
Recent developments have extended conformal prediction to online settings and sequential decision-making applications. Adaptive Conformal Inference (ACI)~\cite{gibbs2021adaptive} enables real-time adaptation to distribution shifts by dynamically adjusting prediction set sizes based on recent coverage performance. Rolling Risk Control~\cite{feldman2023achieving} extends ACI to control arbitrary bounded loss functions. The conformal PID controller~\cite{angelopoulos2023} bridges conformal prediction and control theory, enabling more stable coverage in an online adaptive setting.
Recent work on conformal decision theory~\cite{lekeufack2024} further generalizes these concepts establishing the theoretical foundations for safe autonomous decisions under imperfect predictions.

\yq{
\section{Problem Formulation and Preview of Main Results}\label{sec:Prob}}

\subsection{Storage Arbitrage}
We consider the online arbitrage problem of an automated energy storage unit participating in the real-time electricity market \yq{as a price-taking participant}. At each time step $t \in \mathcal{T}$, where $\mathcal T = \{1,\ldots, T\}$  \yq{defines the arbitrage window}, the unit must decide whether to charge, discharge, or remain idle, with the objective of maximizing the total profit comprised of both the immediate payoff and the expected maximized profit over the remaining arbitrage window \yq{$\{t+1,\ldots, T\}$}.
We denote $Q_t(e_t)$ as the opportunity value function of the battery's SoC $e_t \in [0,E]$ at time step $t$. $Q_t(e_t)$ is monotonically non-decreasing in $e_t$. Note that the true value of $Q_t(e_t)$ is unknown in real time and only revealed retrospectively after the entire arbitrage window, hence we must resort to predictions for online decision-making.
We denote $\hat{Q}_t(e_t|\lambda_t)$ as the \emph{predicted} opportunity value function with respect to SoC~$e_t$ given the realization of price~$\lambda_t$ (the detailed prediction method is described in Section~\ref{sec:forecast}).
The optimal arbitrage problem takes the form:
\allowdisplaybreaks
\begin{subequations}
\begin{align}
\maximize{p_t, b_t, e_t} & \quad \lambda_t (p_t - b_t) - C p_t + \hat{Q}_t(e_t | \lambda_t) \label{eq:arbitrage_obj}\\
  \text{s.t.} & \quad   0 \leq p_t,\ b_t \leq P \label{eq:1b} \\
  & \quad p_t = 0 \quad \text{if} \quad \lambda_t < 0 \label{eq:1c} \\
  & \quad e_t - e_{t-1} = -\frac{p_t}{\eta} + b_t \eta \label{eq:1d} \\
  & \quad 0 \leq e_t \leq E \label{eq:1e}
\end{align}
\label{eq:arbitrage}
\end{subequations}
where the first term in the objective function~\eqref{eq:arbitrage_obj} represents the arbitrage revenue, calculated as the product of the real-time market price $\lambda_t$ and the net power dispatch $(p_t - b_t)$, with~$p_t$ representing discharge power and~$b_t$ representing charge power.
The second term accounts for discharge costs, where $C$ represents the marginal operational cost, including factors such as battery degradation.
The third term $\hat{Q}_t(e_t|\lambda_t)$ is the \emph{predicted} opportunity value to go.
Constraint~\eqref{eq:1b} enforces the power rating limit $P$ for both charging and discharging operations. Constraint~\eqref{eq:1c} provides a relaxed formulation that prevents simultaneous charging and discharging~\cite{xu2020}. \yq{This assumption is reasonable since discharging at negative prices is economically irrational.} Constraint~\eqref{eq:1d} models the SoC evolution constraint with charging/discharging efficiency $\eta$. Constraint~\eqref{eq:1e} ensures the SoC remains within the storage capacity limits.
To ease notation we define~$\mathcal{E}(e_{t-1})$ as the set $(p_t, b_t, e_t)$ such that constraints~\eqref{eq:1b}-\eqref{eq:1e} are feasible given  $e_{t-1}$. 

Let $q_t(e_t)$ denote the marginal opportunity value function:
\begin{align}
    q_t(e_t) := \frac{\partial}{\partial e_t} Q_t(e_t). \label{eq:marginal}
\end{align}
The  predicted marginal opportunity value function is defined analogously using $\hat{q}_t(e_t|\lambda_t)$. \footnote{In the remaining sections, we use $\hat{Q}_t(e_t)$ and $\hat{q}_t(e_t)$ instead when the price input is clear from context.}

By adopting a dynamic programming approach (detailed in Section~\ref{sec:forecast}), it can be seen that marginal opportunity value functions $q_t(e_t)$ and thus $\hat{q}_t(e_t|\lambda_t)$ are non-increasing in $e_t$, which makes the objective function~\eqref{eq:arbitrage_obj} concave in $e_t$. \yw{It is assumed  that the feasible set~$\mathcal{E}(e_{t-1})$ is nonempty.} Therefore, the optimization problem~\eqref{eq:arbitrage} is convex with a concave-maximizing objective function and non-empty feasible set~\cite{boyd2004}. The predicted marginal opportunity value function, $\hat q_t(e_t|\lambda_t)$, crucially appears in the  analytical solutions to~\eqref{eq:arbitrage}, i.e., the storage dispatch decisions~$(p^*_t,b^*_t)$ for all $t \in \mathcal{T}$. These solutions can be obtained with the realized price~$\lambda_t$, the prediction of marginal opportunity value function and current SoC~$e_{t-1}$~\cite{zheng2022}. Detailed formulation and interpretations of the solution are provided in Appendix~\ref{app:arbitrage_solution}. Finally, we denote the dispatch decision at time $t$ as $D_t := (p^*_t,b^*_t)$.

\yq{
\begin{remark}
Storage arbitrage problem~\eqref{eq:arbitrage} applies to both price-taker market bidding and price response modes. 
The key distinction lies in the time of decision-making relative to market price realization.
\begin{enumerate}
    \item  Market bidding: storage submits its bids a predetermined time before the market clearing period $t$ and follows the dispatch results determined by market clearing. Given that bids for time step $t$ are designed indeed based on the price forecast for $\{t+1,\ldots, T\}$ and market clearing relies only on  current bids, the joint problem of optimal bid design and market clearing is equivalent to the formulation~\eqref{eq:arbitrage}~\cite{baker2023transferable,yi2025decision}. 
    \item  Price response: storage updates its dispatch decisions at each time step $t$ after observing the most recent real-time market price $\lambda_t$, announced by the system operator prior to dispatch.
\end{enumerate}

\end{remark}
}

\subsection{Opportunity Value Function Forecasting}\label{sec:forecast}
The prediction of the marginal opportunity value function~\( \hat{q}_t(e_t) \) can be obtained through a forecasting model proposed in~\cite{baker2023transferable}, which we briefly review here.
The development of this forecasting model involves two key steps: opportunity valuation and model training.
First, 
notice that the optimal arbitrage problem~\eqref{eq:arbitrage} can be formulated within a dynamic programming framework,
where the total value given  time step represents the maximum achievable arbitrage profit comprising both the payoff from the current time step (the first two terms in~\eqref{eq:arbitrage_obj}) and the potential opportunity values from future operations (the third term in~\eqref{eq:arbitrage_obj}). Thus, given a series of prices~$\{\lambda_t\}_{t\in \mathcal T}$, this opportunity value function can be calculated in a backwards recursive manner, as follows:
\begin{align}
    Q_{t-1}(e_{t-1}) &= \maximize{\substack{p_t, b_t, e_t  \\\in\mathcal{E}(e_{t-1}) }} \quad \lambda_t (p_t - b_t) - C p_t + Q_t(e_t) \label{eq:dp}
\end{align}
This approach follows the Bellman equation in dynamic programming~\cite{bellman1957dynamic} and fits a piece-wise linear approximation of the marginal opportunity value function $q_t(e_t)$ based on the first-order derivative of the optimal value function $Q_t(e_t)$, where $e_t$ is drawn from the set of SoC segments $\mathcal S$.
The valuation of $q_t(e_t)$ can be efficiently computed using analytical solutions in Appendix~\eqref{eq:control}. Once the opportunity value function is calculated, the forecasting model can be trained using historical (and current) \yq{day-ahead and real-time prices} $\Lambda_t$ over a look-back window and the corresponding marginal opportunity value function $q_t(e_t)$. We denote this model as~$\mathcal{M}:\Lambda\rightarrow \hat{q}$, where $\Lambda$ is the set $\{\Lambda_t\}_{t\in \mathcal T}$ and $\hat{q}$ is $\{\hat{q}_t(e_t|\lambda_t)\}_{t\in \mathcal T}$.

Upon the calculation of the opportunity value function, we'll find a piece-wise linear structure in $q_t(e_t)$ and thus $\hat{q}_t(e_t)$, as illustrated in Figure~\ref{fig:bid}(\subref{fig:bid1}). Such structure exposes the non-uniqueness of optimal solutions to problem~\eqref{eq:arbitrage} when the electricity price $\lambda_t$ coincides with the flat segments of $\hat{q}_t(e_t)$, as shown in Figure~\ref{fig:bid}(\subref{fig:bid2}). To ensure the uniqueness of optimal solutions, we make the following assumption:

\begin{assumption}[Non-Degenerate Prices]\label{assump:unique}
For any time step $t \in \mathcal{T}$, the event $\{\lambda_t \in \mathcal{S}_t\}$ occurs with probability zero, where $\mathcal{S}_t$ denotes the set of constant values corresponding to flat segments of $\hat{q}_t(e_t)$, i.e., $\mathbb{P}(\lambda_t \in \mathcal{S}_t) = 0$.
\end{assumption}    

Assumption~\ref{assump:unique} is reasonable for storage arbitrage since electricity prices are typically continuous random variables, thus they almost never take specific threshold values. This ensures unique optimal solutions to problem~\eqref{eq:arbitrage} under normal market conditions.

\begin{figure*}[!t]
    \centering
    \includegraphics[width=.95\linewidth]{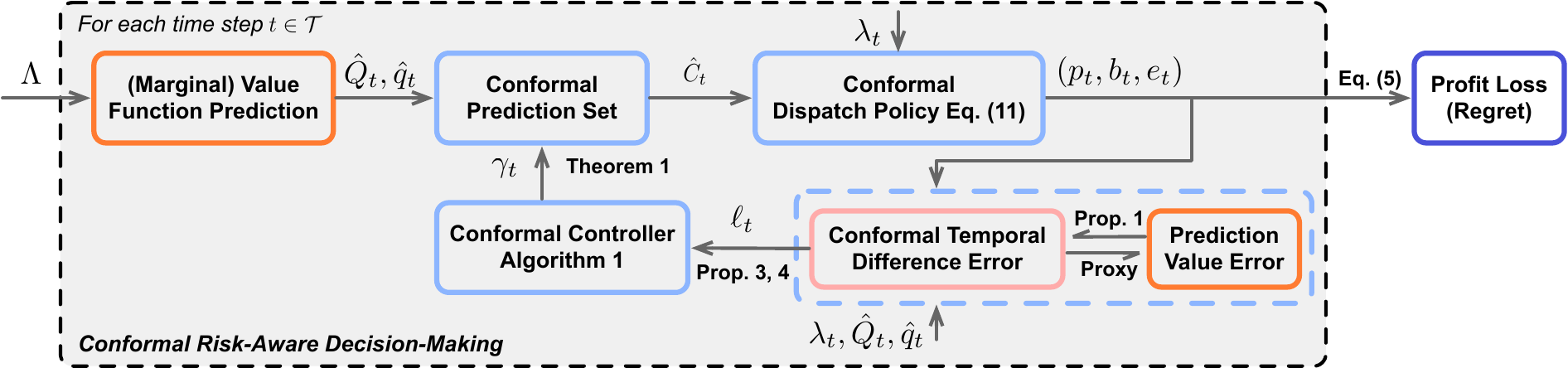}
    \caption{A flowchart of the proposed conformal risk-aware energy storage arbitrage strategy.}
    \label{fig:logic}
\end{figure*}

\subsection{Decision Evaluation under Uncertainty}

Policy evaluation for online decision-making is non-trivial since decisions are made under uncertainty about future states, making the true quality of current actions unobservable until the outcomes fully unfold.
In the context of storage arbitrage, the main objective of  policy evaluation is to quantify the (prediction) value error: 
\begin{equation}
    \Delta(e_t) := \hat{Q}_t(e_t|\lambda_t) - Q_t(e_t) \label{eq:value_error}
\end{equation}
which represents the distance from the forecast to the true value function. However,  $\Delta(e_t)$ is often unmeasurable as the true value function $Q_t(e_t)$ is unobtainable in real time.
To analyze this $\Delta_t$, we first reformulate the opportunity valuation~\eqref{eq:dp} into an online framework as follows:
\begin{align}
\bar{Q}_{t-1}(e_{t-1}|\lambda_t) = \maximize{\substack{p_t, b_t, e_t  \\\in\mathcal{E}(e_{t-1}) }} \quad \lambda_t (p_t - b_t) - C p_t + \hat Q_t(e_t|\lambda_t) \label{eq:Qhat}
\end{align}
where $\bar{Q}_{t-1}(e_{t-1}|\lambda_t)$ denotes the \emph{corrected} opportunity value function at time step $t$ given the price $\lambda_t$ and the current SoC~$e_{t-1}$, with $\bar{q}_{t-1}(e_{t-1}|\lambda_t)$ as its corresponding derivative, defined analogously to~\eqref{eq:marginal}. Note that $\bar{Q}_{t-1}(e_{t-1}|\lambda_t)$ is equal to the objective function value obtained by solving~\eqref{eq:arbitrage}. 
The quantity $\bar{Q}_{t-1}(e_{t-1}|\lambda_t)$ provides an unbiased estimate for~$\hat{Q}_{t-1}(e_{t-1}|\lambda_{t-1})$.

An alternative approximation for value error is the Bellman error~\cite{sutton1998reinforcement}. Under deterministic environment and policy conditions, as in our setup with transition dynamics and policy defined by~\eqref{eq:Qhat}, this Bellman error is identical to the temporal difference error~\cite{sutton1988}, defined as:
\begin{align}
\label{eq:temerr}
    \delta(e_t) := \hat{Q}_t(e_t|\lambda_{t}) - \bar{Q}_{t}(e_{t}|\lambda_{t+1}).
\end{align}

\yw{The Bellman equation has a unique solution given by the true value function~\cite{bertsekas1996}. 
 Alternatively, the solution of the Bellman equation can be viewed from a  temporal-difference perspective in that it is analogous to the temporal difference error being zero~\cite{fujimoto2022}, which will later be formalized in Proposition~\ref{prop:bellman}.} 
Due to the existence of such errors, sequential online decisions under opportunity value function forecasting inevitably yield suboptimal results with respect to perfect information.
In the context of storage arbitrage, this value error manifests as a \emph{profit loss} relative to the theoretical optimum, as further discussed in \yq{Section~\ref{sec:problem_statement}}.

\begin{remark}
    Although the discussion on decision evaluation under uncertainty above is framed based on prediction errors in the value function~$\hat Q_t(e_t)$, similar results hold for errors in the marginal value function~$\hat{q}_{t}(e_{t})$.
\end{remark}

\subsection{Problem Statement}\label{sec:problem_statement}

    Following Bellman's principle of optimality~\cite{bellman1957dynamic}, the solution to the arbitrage problem~\eqref{eq:dp} yields the optimal storage dispatch decisions that maximize the total profit over the arbitrage window. That is, when opportunity value functions are estimated accurately, i.e., $\hat{Q}_t(e_t)={Q}_t(e_t)$  for all $e_t \in [0,E], \ t \in \mathcal T$, the calculated per-step decisions are globally optimal. In this ideal case, it's easy to show that the total value, i.e., incurred profits plus remaining opportunity value, remains constant throughout the arbitrage window. A strategy achieving this is said to fully capture and realize the theoretical opportunity value. The ultimate objective of a risk-aware arbitrage strategy is to constrain the gap between the theoretical optimum and realized discounted profits, i.e., \emph{profit loss} in hindsight. Such a gap is often referred to as \emph{regret} in the online algorithm literature. 
    \yq{Let $\Pi_t$ denote the \emph{step arbitrage profit} as the sum of first two terms in~\eqref{eq:arbitrage_obj}. }
    Given a certain sequence of decision decisions (dispatch policy)~$\mathcal D_{1:T}:=\{(p_{t}, b_{t})\}_{t\in \mathcal T}$, we formally define this as follows: 
    \begin{align}
        \text{Regret}: =  \max_{\substack{{p}, {b}, {e}  \in {\mathcal{E}} }} \sum_{t=1}^T{\Pi_t}({p}, {b}) - \sum_{t=1}^T{\Pi_t}({p}, {b}|\mathcal{D}_{1:T}).
        \label{eq:regret}
    \end{align}

Our goal is to provide a risk-aware approach that calibrates online arbitrage decision-making to manage the risk exposure stemming from the imperfect (marginal) value function prediction error. This calibration leverages available online information and indirectly limits downside risk in profit by constraining the direct forecasting impacts within a predefined bound.

\yq{
\subsection{Proposed Approach Overview}\label{sec:overview}}
A flowchart of the proposed approach is summarized in Figure~\ref{fig:logic}. First, as in
\yq{Section~\ref{sec:problem_statement}},
 we establish that imperfect predictions inevitably lead to suboptimal decisions and potential profit losses for energy storage arbitrage. Though such profit losses are inherently unobservable in real-time operations, they can be estimated using the temporal difference error as a proxy \yq{as later demonstrated in Proposition~\ref{prop:bellman}}. Therefore, the key insight underlying our proposed approach is to constrain the arbitrage \emph{cumulative risk}, quantified through the observable temporal difference error (referred to as the \emph{loss of interest} in the following sections). Controlling this cumulative risk then serves as an indirect mechanism for managing the unobservable profit risk. This risk management approach is achieved through a conformal decision-making framework that dynamically calibrates a conformal control variable, with detailed methodology presented in Section~\ref{sec:Meth} and the main theoretical results in Theorem~\ref{theorem:cc}.
The online calibration framework supports various practical loss metrics implementable in real-time decision-making, such as temporal prediction error, temporal value error, etc. A comprehensive discussion of these loss functions and their practical implications is provided in Section~\ref{sec:loss}.

This framework shares similarities with regret analyses, as it retrospectively evaluates decision quality against optimal benchmarks. However, regret analyses typically apply to settings where the loss of interest is measurable after the implementation of decisions, while in our case, the loss is unobservable and must be estimated based on the available information.

\section{Conformal Risk-Aware Energy Storage Arbitrage}\label{sec:Meth}
The arbitrage problem~\eqref{eq:arbitrage} describes a \emph{risk-neutral} framework. We now consider an arbitrage strategy in a \emph{risk-aware} context that mitigates downside risk, utilizing the conformal decision theory~\cite{lekeufack2024}. 
In this section, we first briefly summarize the main results of this theoretical framework and then establish its compatibility in the context of our problem.

\subsection{Conformal Decision Theory}

Conformal decision theory provides a methodology for calibrating energy storage's decisions to achieve statistical guarantees on realized average losses of interest, i.e., \emph{cumulative risk}.  
This framework can be formulated in a more abstract and notationally convenient manner as follows. 
For any sequence of families of dispatch decisions $\mathcal{D}_{1:T}$ that are \emph{eventually safe} (defined formally in Definition~\ref{def:1}, with high-level idea here), the theory establishes that they can be calibrated online via a \emph{conformal control variable}~$\gamma \in \mathbb{R}$ to achieve bounded long-term risk. The calibration goal is formulated as:
\begin{align}
\text{find }  \gamma_{1:T} \quad \text{s.t. } \hat{R}_T(D_{1:T}, \gamma_{1:T}) \leq \varepsilon + \frac{a \cdot \phi(T)}{T} \label{eq:risk_bound}
\end{align}
where $\varepsilon \in [0, 1]$ is a predefined risk tolerance level, $a$ is a (small) constant, and $\phi(T)$ is any sublinear function satisfying  $\lim_{T \to \infty} \phi(T)/T = 0$. The cumulative risk is defined as:
\begin{align}
\hat{R}_T(D_{1:T}, \gamma_{1:T}) := \frac{1}{T} \sum_{t=1}^{T} \mathcal{L}(D_t^{\gamma_t}(x_t), y_t)  \text{ and } \hat{R}_0 = 0. \label{eq:cum_risk}
\end{align}
where state-value pairs are defined as
\begin{align}
x_t = \big(e_{t-1}, \lambda_t, \hat{Q}_t(e_t|\lambda_t)\big) ~~ \text{and} ~~ y_t =\bar{Q}_t(e_t|\lambda_{t+1})\text{ or }\bar q_t(e_t|\lambda_{t+1}) \label{eq:xy}
\end{align}
respectively.

The dispatch decision  $D_t^{\gamma_t}$ is parameterized by $\gamma_t$, which controls the level of conservativeness with higher values of $\gamma_t$ corresponding to more aggressive strategies (as specified  later in Section~\ref{sec:cc_arbitrage}).
We evaluate the \emph{quality of the decision-making} through a \emph{loss function} $\mathcal{L}(\cdot)$.\footnote{The framework works for any bounded loss; however, to simplify analysis, we assume the loss is normalized to the interval [0, 1].} Typically, the potential loss increases when aggressive decisions are taken (large~$\gamma_t$), while overly small choices of $\gamma_t$ leads to conservative and under-performing decisions. This trade-off often results in \emph{monotone or near-monotone loss functions} with respect to $\gamma$.

\begin{definition}[Eventually Safe Dispatch]\label{def:1}
    In the storage arbitrage setting, we say that a dispatch policy $\mathcal{D}_{1:T}$ is \emph{eventually safe} if there exists threshold $\varepsilon^{\text{safe}} \in [0,1]$, $\gamma^{\text{safe}} \in \mathbb{R}$, and a horizon $K > 0$ such that for any time step $[t, t+K-1] \subseteq [1,T]$ and any sequence of state-value pairs $\{(x_k, y_k)\}_{k=t}^{t+K-1}$:
\begin{align*}
&\text{If } \lambda_k \leq \lambda^{\text{safe}} \text{ for all } k \in [t, t+K-1], \\
&\qquad \qquad \text{ then } \frac{1}{K} \sum_{k=t}^{t+K-1} \mathcal{L}(D_k^{\lambda_k}(x_k), y_k) \leq \varepsilon^{\text{safe}}.
\end{align*}
\end{definition}

For storage arbitrage, Definition~\ref{def:1} means that 
the realized losses of interest are kept within an acceptable tolerance after a finite number of time steps. In the case of using \emph{prediction sets} to calibrate dispatch decisions, this guarantee is trivial, since one could always ensure profit loss safety by committing to the entire feasible action space, effectively avoiding constraint violations but yielding no economic value.

Conformal controllers~\cite{lekeufack2024} offer a simple yet effective way to address this challenge for storage arbitrage. They calibrate dispatch decisions by adjusting the control parameter $\gamma_t$ to maintain statistical guarantees on realized losses of interest under uncertainty. Let $\mathcal T^+$ denote the set of all time steps $t \in \{1,\ldots,T+1\}$.  
Building on conformal controller design from~\cite[Theorem~1]{lekeufack2024}, we extend its context to energy storage arbitrage and present the following conformal controller. The proof is also adapted accordingly and revised for  clarity in Appendix~\ref{app:theorem_cc}.

\begin{theorem}[Conformal Controller for Storage Arbitrage]\label{theorem:cc}
  Consider the following update rule for the storage arbitrage risk control variable $\gamma_{1:T}$ in~\eqref{eq:cum_risk}:
\begin{align}
\gamma_{t+1} = \gamma_t + \rho \left( \varepsilon - \ell_t \right), \quad \forall t \in \mathcal T
\label{eq:cc}
\end{align}
where $\rho > 0$ and $\ell_t := \mathcal{L}(\mathcal{D}^{\gamma_t}_t(x_t), y_t)$ denotes the realized loss of interest at time step $t$, evaluated under the storage dispatch decision $D_t^{\gamma_t}$ given state-value pair $(x_t,y_t)$, where $x_t$ and $y_t$ are defined by~\eqref{eq:xy}.

If the dispatch sequence $\mathcal{D}_{1:T}$ is eventually safe (Definition~\ref{def:1}) for some horizon $K \in \mathcal T$ with tolerance $\varepsilon^{\text{safe}} \leq \varepsilon$, then for any realization of electricity prices, the arbitrage cumulative risk is bounded:
\begin{align}
\hat{R}_t(D_{1:t}, \gamma_{1:t}) \;\leq\; \varepsilon \;+\; O(1/t),
\label{eq:risk_bound_cc}
\end{align}
for all $t \in \{K, \ldots, T\}$.
\end{theorem}

Theorem~\ref{theorem:cc} establishes the boundedness of cumulative risk for energy storage online arbitrage. It is important to emphasize that this bounded risk does not directly correspond to profit loss, but rather serves as a measurable and effective proxy for it, \yq{as formalized below.}

\yq{
  \begin{proposition}\label{prop:bellman}
    Consider a finite horizon~$\mathcal T$, assuming $\Delta(e_T) = 0$ for all $e_T \in [0,E]$. Under Assumption~\ref{assump:unique}, the temporal difference error is zero, i.e., $\delta(e_t) = 0$, if and only if the value error $\Delta(e_t) = 0$, for all $e_t \in [0,E]$ for all $t \in \mathcal T$.
\end{proposition}

The proof of Proposition~\ref{prop:bellman} is provided in Appendix~\ref{app:bellman}. 
This result suggests that the temporal difference error serves as a measurable proxy for value error, which is the true but typically unobservable objective.
Here, this means the arbitrage cumulative risk serves as a proxy for the underlying profit loss risk.}
We'll explore this relationship further in the following sections.

\begin{remark}

    Though the conformal controller~\eqref{eq:cc} defined in Theorem~\ref{theorem:cc} doesn't explicitly require any monotonicity in the storage arbitrage loss function~$\ell_t$, this strategy becomes practically effective only if the loss is monotone or near-monotone in the conformal control variable~$\gamma_t$.
    Otherwise, the assumption on eventual safety in Definition~\ref{def:1} will not hold.
    More details on conformal risk control and the monotonicity of loss functions can be found in~\cite{angelopoulos2023risk}.
    The monotonicity of the loss functions in our storage arbitrage algorithm will be discussed in more detail in Section~\ref{sec:loss},\yw{where we introduce two choices of loss function}.
\end{remark}

\subsection{Risk-Aware Arbitrage using Conformal Decision Theory}\label{sec:cc_arbitrage}

We now deploy the conformal controller~\eqref{eq:cc} to address uncertainties and establish risk control in our arbitrage framework. To implement this methodology, we construct a conformal prediction set $\hat{C}_t(\hat{q}_t,\gamma_{t-1})$ at each time step $t$, \yw{which is  centered around} the marginal opportunity value function forecasts~$\hat{q}_t$. \yw{This set bounds the uncertainty in forecasting~$\hat{q}_t$} through a specified conformal control variable~$\gamma_{t-1}$ as follows:
\begin{align}
    \hat{C}_t (\hat{q}_t,\gamma_{t-1}) := [\hat{q}_t - \sigma z_{1-\gamma_{t-1}/2}, \hat{q}_t + \sigma  z_{1-\gamma_{t-1}/2}]
    \label{eq:cc_set}
\end{align}
where $\sigma$ is a fixed sensitivity parameter, $z_\gamma$ represents the quantile of level $\gamma$ of the normal distribution. 
The choice of $\sigma$ directly correlates with market volatility, enabling flexibility across different market conditions. \yw{At this point it is worth some time to relate the prediction set $\hat C_t$ to the accuracy of the forecasting model $\mathcal M$. If $\mathcal M$ was perfectly accurate then $\hat C_t$ would simply be the singleton $\hat q_t$. Not only is perfect forecasting not possible, but so is our ability to accurately quantify our forecasting error. Assume that $\epsilon_t$ is the ground truth forecast accuracy, where $\epsilon_t=0$ denotes a perfect forecast, in this setting we would have $\hat{C}_t = \hat{q}_t$. As $\epsilon_t$ increases, i.e., the forecast quality decreases, then $\hat C_t$ will also increase. As we cannot know $\epsilon_t$, we instead parametrize the prediction set by a proxy, $\gamma_t$, that reflects our confidence in the interval. We make the relationship between $\epsilon_t$ and $\gamma_t$ precise later in Proposition~\ref{prop:optimal_width} and provide an illustrative example in Section~\ref{sec:example}.}

The conformal control variable $\gamma_t$ controls the width of the conformal prediction set. A wider interval reflects lower confidence in the prediction and yields a more conservative dispatch policy. The update of the conformal control variable depends on the definition of loss function, which will be specified in the next section.

The conformal dispatch policy $\{D_t^{\gamma_t}\}_{t \in \mathcal{T}}$ applies the conformal prediction set~\eqref{eq:cc_set} to the marginal opportunity value function forecasts~$\hat{q}_t$ for each time step $t$ in the optimal arbitrage problem~\eqref{eq:arbitrage}. The risk-neutral dispatch policy \yw{derived in~\cite{zheng2022} and described in Appendix~\eqref{eq:control}} then becomes:
\begin{subequations}
\begin{align}
    p^*_t & = \min\{\hat{p}_t^{\gamma}, e_{t-1} \eta\}\label{eq:decision_cap_cc}\\
    b^*_t & = \min\{\hat{b}_t^{\gamma}, (E - e_{t-1})/\eta\} \label{eq:decision_cap_b_cc}
\end{align}
where
{\small
\begin{align}
  (\hat{p}_t, \hat{b}_t)^{\gamma} = 
  \begin{cases} 
  (0, P) & \text{if } \lambda_t \leq \min\{\hat{C}_t(e_{t-1} + P \eta)\} \eta \\
  \{0, \alpha^{\gamma}\} & \text{if } \min\{\hat{C}_t(e_{t-1} + P \eta)\} \eta < \lambda_t \\
  &\qquad\leq \min\{\hat{C}_t(e_{t-1})\} \eta\\
  (0, 0) & \text{if } \min\{\hat{C}_t(e_{t-1})\} \eta < \lambda_t \\
&\qquad\leq [\max\{\hat{C}_t(e_{t-1})\}/\eta + C]^+ \\
  (\beta^{\gamma}, 0) & \text{if } [\max\{\hat{C}_t(e_{t-1})\}/\eta + C]^+  < \lambda_t \\
  &\qquad\leq [\max\{\hat{C}_t(e_{t-1} - P/\eta)\}/\eta + C]^+ \\
  (P, 0) & \text{if } \lambda_t > [\max\{\hat{C}_t(e_{t-1} - P/\eta)\}/\eta + C]^+
  \end{cases}
\end{align}}
and
\begin{align}
\alpha^{\gamma} &= \frac{\hat{q}_t^{-1}\left(\lambda_t/\eta + \sigma  z_{1-\gamma/2}\right) - e_{t-1}}{\eta}\\
\beta^{\gamma} &= \left(e_{t-1} - \hat{q}_t^{-1}\left(\left(\lambda_t - C \right)/ \eta - \sigma  z_{1-\gamma/2}\right)\right){\eta}
\end{align}
\label{eq:control_cc}
\end{subequations}

\begin{figure}[t]
    \centering
    \begin{subfigure}[b]{0.48\textwidth}
        \centering
        \includegraphics[width=\textwidth]{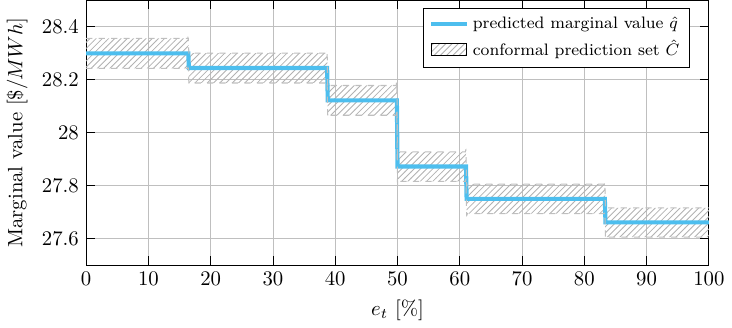}
        \caption{SoC-dependent marginal opportunity value function.}
        \label{fig:bid1}
    \end{subfigure}
    \hfill
    \begin{subfigure}[b]{0.48\textwidth}
        \centering
        \includegraphics[width=\textwidth]{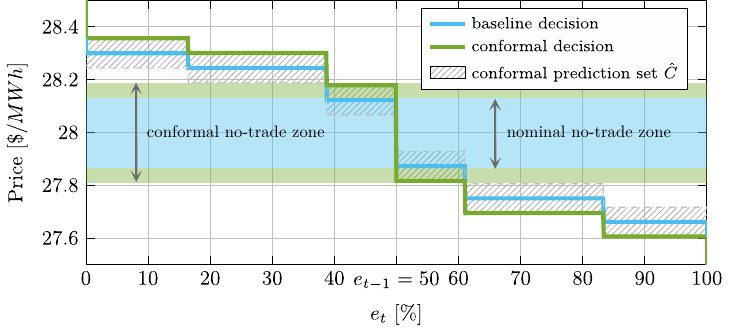}
        \caption{SoC-dependent dispatch decisions.}
        \label{fig:bid2}
    \end{subfigure}
    \caption{Marginal opportunity valuation and energy storage arbitrage decision-making using conformal prediction set \yw{$P = 0.5 MW$,~$E = 1.0 MWh$}}.
    \label{fig:bid}
  \end{figure}

The interpretation of~\eqref{eq:control_cc} follows the same logic as the optimal solutions to the arbitrage problem~\eqref{eq:arbitrage}, which is detailed in~\eqref{eq:control} in Appendix~\ref{app:arbitrage_solution}, with the key difference being the use of the conformal prediction sets~\eqref{eq:cc_set}. The impact of such prediction set is visualized in Figure~\ref{fig:bid}. In this example, we assume power rating~$P = 0.5 MW$ and capacity~$E = 1.0 MWh$.
Figure~\ref{fig:bid}(\subref{fig:bid1}) displays the marginal opportunity value function (blue line), a non-increasing piece-wise linear function of SoC~$e_t$.
The shaded area represents the conformal prediction set~$\hat{C}_t$ defined in~\eqref{eq:cc_set}.
Figure~\ref{fig:bid}(\subref{fig:bid2}) illustrates how the unit makes dispatch decisions based on the current price~$\lambda_t$, initial SoC~$e_{t-1}$ (assumed~$e_{t-1} = 50\%$), and operational conditions~$\mathcal{E}(e_{t-1})$.
The blue line represents baseline decisions directly adopting the predicted opportunity value function from Figure~\ref{fig:bid}(\subref{fig:bid1}). The green line represents decisions adjusted by the conformal prediction set.
Both these two sets of decisions share the following principle: the left-hand side of the initial SOC means the unit is discharging given the current price $\lambda_t$, while the right-hand side means the unit is charging; there is a no-trade zone (price spread) around the initial SoC due to operational costs and efficiency losses; and operations are bounded by power rating and storage capacity limits regardless of price extremes.

In particular, the strategy aided by conformal prediction set exhibits several key characteristics: 
\begin{itemize}
    \item Establishes a wider no-trade zone, as price movements within the prediction set remain uncertain, requiring a larger price spread to compensate for this uncertainty;
    \item Discharges only when prices are high with high confidence; and
    \item Charges only when prices fall below the prediction set where future price increases are likely.
\end{itemize}

This strategy makes the unit more selective about trading opportunities, reflecting a risk-aware preference. \yq{For example, using a smaller $\gamma$ in the conformal strategy creates wider prediction sets, which reduces trading frequency. Though fewer trades are taken, the expected profit per trade increases, resulting in more conservative operation but higher confidence in profitability. As a result, decision-making becomes less sensitive to the underlying value function forecast differences.}

\begin{algorithm}[t]
    \caption{Arbitrage using Conformal Controller}
    \label{alg:arbitrage}
    \begin{flushleft}
    \textbf{Input:}\\
    \begingroup
    \leftskip=1em
    \rightskip=0em
    Setup information $\{\{\lambda_t\}_{t\in \mathcal T}, C, P, \eta, E\}$, forecasting model $\mathcal{M}:\Lambda\rightarrow \hat{q}$, controller parameters $\{\varepsilon \in \mathbb{R}, \rho > 0, \sigma\}$.
    \par
    \endgroup
    \textbf{Process:}
    \begin{algorithmic}[1]
    \STATE  Initialize $e_0$ and conformal variable $\gamma_0$.
    \STATE  \textbf{for} $t = 1, \ldots, T$ \textbf{do}
    \STATE  \hspace*{1em}Obtain $\lambda_t$ and construct forecaster input $\Lambda_t$.
    \STATE  \hspace*{1em}Predict $\hat{q}_t(e_t)\leftarrow\mathcal{M}(\Lambda_t)$.    
    \STATE  \hspace*{1em}Construct a prediction set: $\hat{C}_t(\hat{q}_t, \gamma_{t-1})$.
    \STATE  \hspace*{1em}Compute decisions $(p^*_t, b^*_t)$ based on \eqref{eq:control_cc}.
    \STATE  \hspace*{1em}Update $e_t \leftarrow e_{t-1}- p^*_t / \eta + b^*_t \eta$.
    \STATE  \hspace*{1em}\yw{Compute $\ell_t \leftarrow \mathrm{\texttt{clip}}(\ell_t)$ according to~\eqref{eq:tloss1} or~\eqref{eq:tloss2}.}\label{line:loss}
    \STATE  \hspace*{1em}Update $\gamma_t \leftarrow \gamma_{t-1} + \rho(\varepsilon - \ell_t)$. \label{line:gamma}
    \STATE  \textbf{end for}
    \end{algorithmic}
    \textbf{Output:}\\
    \begingroup
    \leftskip=1em
    \rightskip=0em
    dispatch decisions $(p^*_t, b^*_t)$ for each time step $t \in \mathcal T$.
    \par
    \endgroup
    \end{flushleft}
    \end{algorithm}

\yq{

\begin{proposition}\label{prop:optimal_width}
    Given a fixed forecast accuracy level $\epsilon_t$ (where smaller values indicate better accuracy), 
    there exists (at least) one corresponding conformal control variable $\gamma^*_t$ such that the prediction set $\hat C_{t+1}$ given by~\eqref{eq:cc_set} and the resulting dispatch policy~\eqref{eq:control_cc} maximizes the total profit, and the value of such $\gamma^*_t$ is non-increasing in $\epsilon_t$ in expectation.
\end{proposition}

The proof of Proposition~\ref{prop:optimal_width} is provided in Appendix~\ref{app:optimal_width}. 
This proposition establishes 
that each level of (observed) forecast accuracy $\epsilon_t$ corresponds to an appropriate conformal control variable $\gamma_t$, and thus an appropriate prediction set width, as in~\eqref{eq:cc_set}, allowing the storage to optimally balance risk and opportunity.

}

    The proposed risk-aware arbitrage strategy using the conformal controller is summarized in Algorithm~\ref{alg:arbitrage}.    
    An illustrative example regarding arbitrage performance is provided in the following subsection, where we compare the conformal risk-aware strategy with a risk-neutral strategy.

\subsection{Illustrative Example of Conformal Risk-Aware Arbitrage}\label{sec:example}

\begin{figure}[t]
    \centering
    \begin{subfigure}[b]{0.48\textwidth}
        \centering
        \includegraphics[width=\textwidth]{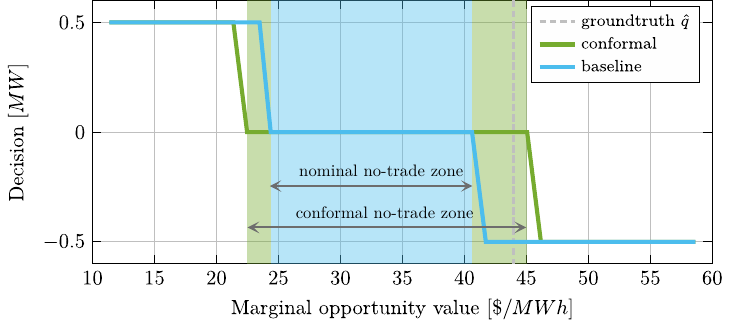}
        \caption{Dispatch decision.}
        \label{fig:tloss1}
    \end{subfigure}
    \hfill
    \begin{subfigure}[b]{0.48\textwidth}
        \centering
        \includegraphics[width=\textwidth]{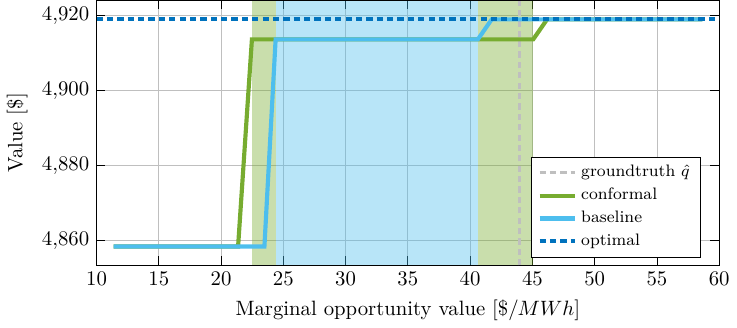}
        \caption{Total value.}
        \label{fig:tloss2}
    \end{subfigure}
    \caption{Single time step decision-making and value analysis given various marginal opportunity value function forecast accuracy ($P = 0.5 MW$,~$E = 1.0 MWh$). }
    \label{fig:tloss}
\end{figure}

Without loss of generality, we can assume that the marginal opportunity value function converges to a single value at each time step $t$, i.e., a single-segment marginal value function. Figure~\ref{fig:tloss} illustrates a single time step snapshot of decision-making and value analysis under this setup, given of varying marginal opportunity value function forecast accuracy. As before, we assume the unit operates with a setup with power rating~$P = 0.5 MW$ and capacity~$E = 1.0 MWh$, the conformal strategy relies on a fixed $10\%$ wide prediction set width, and the optimal decision in this case would be for the unit to fully charge. The gray vertical line in Figures~\ref{fig:tloss}(\subref{fig:tloss1}) and~\ref{fig:tloss}(\subref{fig:tloss2}) gives the groundtruth marginal opportunity value function, while the x-axis represents the predicted values. 

Figure~\ref{fig:tloss}(\subref{fig:tloss1}) shows the dispatch decisions based on the forecast of marginal value function, where the blue line represents the decisions made directly with the forecast, i.e., risk-neutral strategy with~\eqref{eq:control}, and the green line shows the risk-aware conformal strategy using~\eqref{eq:control_cc}. Here the y-axis represents the net power dispatch $(p_t - b_t)$, e.g., a value of~$0.5MW$ indicates discharge at full capacity.
With the vanilla strategy, when the forecast is accurate or higher than the groundtruth, the unit would also decide to charge at full capacity.
However, if the forecast is lower than the groundtruth, the unit would make alternative decisions to stand by or even discharge if it wrongly predicts an extremely low future value. In contrast, the conformal strategy is more conservative, extending the idle period with respect to the forecast. It only discharges when the forecast is significantly lower than the groundtruth, and vice versa for charging decisions. This creates two forecast ranges where decisions from the two strategies differ. 

We provide an ex-post value analysis in Figure~\ref{fig:tloss}(\subref{fig:tloss2}), comparing the total value after applying the corresponding decisions corrected by the groundtruth opportunity values. First, any decisions deviating from those made with accurate forecasts lead to suboptimal value, i.e., value below the offline optimal value (dark blue line). 
Second, the difference in profit loss between the two strategies shows different patterns in two forecast ranges: within~$\$21/MWh-\$24/MWh$, the conformal strategy yields a lower profit loss by avoiding exposure to significantly incorrect forecasts; within $\$41/MWh-\$46/MWh$ (closer to the groundtruth), the conformal strategy yields a higher profit loss than the vanilla strategy as it is conservative in making decisions. Overall, when forecasts are fairly accurate, adopting a conformal prediction set may lead to higher profit loss; however, when forecasts are inaccurate, the unit benefits through mitigated risk of adverse outcomes.

The observation above can be easily extended to other cases where the unit would decide to partially charge or discharge, fully discharge, or stand by. The principle holds that when the forecast is accurate, the unit can be more aggressive in making decisions, i.e., with a narrower prediction set interval; and conversely, when forecast accuracy decreases, a wider prediction set is suggested. 

\section{Online Decision-Making Calibration}\label{sec:loss}
\yw{The example in Section~\ref{sec:example} demonstrated the conformal risk-aware arbitrage strategy for a single time step. We now consider the problem over a time horizon $\mathcal T$ and  introduce an online calibration mechanism that dynamically controls the prediction set width through the conformal control variable $\gamma_t$ as defined in~\eqref{eq:cc_set} and implemented in Line~\ref{line:gamma} in Algorithm~\ref{alg:arbitrage}.
This calibration process adjusts the aggressiveness level of the arbitrage strategy in response to deviations between predicted and realized opportunity values, thereby managing 
the cumulative risk exposure.}
To do so, we utilize temporal difference error as an effective proxy for value error in the loss function design, i.e., Line~\ref{line:loss} in Algorithm~\ref{alg:arbitrage}. Such loss functions serve as the loss of interest, i.e., a temporal loss, for the conformal controller. We propose two distinct strategies: one based on the marginal opportunity value function, and the second based on the opportunity value function respectively.
\subsection{Strategy Calibration via Prediction Error}

Profit loss is essentially determined by the forecast error in the marginal opportunity value function. Here we define a \emph{temporal prediction error} to quantify the distance between the original forecasting and the corrected value. The conformal strategy calibrated by this prediction error is referred to as \emph{CC--prediction err} \yw{(where CC stands for conformal controller)}.

\begin{definition}[\textbf{Temporal Prediction Error}] 
    The (averaged) absolute difference between the corrected marginal opportunity value function forecasting~$\bar q_t(e^*_t|\lambda_{t+1})$ and the executed forecasting~$\hat q_t(e^*_t|\lambda_t)$ is
    \begin{align*}
        \Delta q_{t} = \frac 1 t \sum_{\tau = 1}^t |\hat{q}_{\tau}(e^*_{\tau}|\lambda_{\tau})-\bar{q}_{\tau}(e^*_{\tau}|\lambda_{\tau+1})|,
    \end{align*}
    where $\bar q_t(\cdot)$ is only available after the realization of $\lambda_{t+1}$ as defined in~\eqref{eq:Qhat}, and $e^*_t$ is the executed SoC achieved by taking action $(p^*_t, b^*_t)$.
\end{definition}

To enable online adaptation of the prediction set, here we define a \emph{conservativeness calibration function} $f$ that quantifies the relationship between forecast accuracy and optimal prediction set size, as established in Proposition~\ref{prop:optimal_width}. Using temporal prediction error to represent the forecast accuracy, we can construct $f:\Delta q \rightarrow \hat{\gamma}$, where $\hat{\gamma}$ indicates a reference conformal control variable, and $\hat{\gamma}$ is non-increasing in $\Delta q$. This function can be obtained heuristically from historical data. \yw{One example of such a function is $f:\hat{\gamma}_t \leftarrow \bar{\gamma}  \left(1 -\exp(-k  \Delta q)\right)$, where~$\bar{\gamma}$ represents a maximum threshold of the conformal control variable, $k>0$ is a scaling factor. This form provides a smooth and bounded mapping between the prediction discrepancy~$\Delta q$ and the strategy aggressiveness.
It ensures that $\hat{\gamma}_t \in [0, \bar{\gamma}]$, increasing monotonically with~$\Delta q$
while asymptotically saturating at the maximum threshold~$\bar{\gamma}$.
The scaling factor $k$ controls the sensitivity of this adaptation where larger $k$ leads to faster adjustment, whereas smaller $k$ yields smoother calibration dynamics.} Then the step loss in~\eqref{eq:cc} is calculated as:
\begin{align} 
    \ell_{t+1} = \gamma_t - \hat{\gamma}_t
    \label{eq:tloss1}
\end{align}
Here the time index of step loss is $t+1$ rather than $t$ as $\Delta q_{t}$ (or equivalently, $\bar q_t$) is updated at time step $t+1$, similarly for~\eqref{eq:tloss2} which will be introduced in the next subsection.

\begin{proposition}
    The temporal prediction error-based step loss $\ell_{t+1}$ defined in~\eqref{eq:tloss1} is monotonically non-decreasing in the conformal control variable~$\gamma_t$.
    \label{prop:tloss1}
\end{proposition}

The proof of Proposition~\ref{prop:tloss1} is obvious from definition~\eqref{eq:tloss1}. This calibration controller aims to drive $\gamma_t$
to converge to a level that heuristically maximizes total profit under imperfect forecasts, as suggested by Proposition~\ref{prop:optimal_width}. Note that the effectiveness of this calibration approach depends highly on accurately estimating the mapping function~$f$.

\subsection{Strategy Calibration via Value Error}

Based on the definition of temporal difference error in value functions~\eqref{eq:temerr}, decisions derived from the corrected opportunity value function~$\bar{Q}_t(e_t|\lambda_{t+1})$ and executed policies~$\hat{Q}_t(e_t|\lambda_t)$ fundamentally represents two distinct decision policies.
Performance difference of these two policies quantifies the value error due to the forecast error.
Here we define a \emph{conformal temporal value error} to measure this difference, which is the temporal difference error manifested in the value function after the conformal controller has been deployed, rather than the raw forecast error in the marginal value function.
The conformal strategy calibrated by this value error is referred to as \emph{CC--value err}.
\begin{definition}[\textbf{Temporal Value Error}]
    The difference in value functions by deploying the corrected value function forecasting~$\bar{Q}_t(e_t|\lambda_{t+1})$ and executed forecasting~$\hat{Q}_t(e_t|\lambda_t)$ is denoted as:
    \begin{subequations}
    \begin{align}
        \Delta V_{t} = \bar V_{t}(e_{t-1}) - \hat{V}_{t}(e_{t-1}) \label{eq:Vloss2}
    \end{align}
    where 
    \begin{align}
        \bar V_{t}(e_{t-1})& = \lambda_{t}(\bar p_{t} - \bar b_{t}) - C \bar p_{t} + \bar{Q}_t(\bar e_t) \label{eq:value1}\\
        \hat{V}_{t}(e_{t-1}) &= \lambda_{t}( p^*_{t} -  b^*_{t}) - C p^*_{t} + \bar{Q}_t(e^*_t)\label{eq:value2}
    \end{align}
    Here, $\bar{(\cdot)}$ at the right hand side of~\eqref{eq:value1} and~\eqref{eq:value2} represents the corrected decisions based on~$\bar{q}_t$ (or equivalently,~$\bar{Q}_t$), and~$(\cdot)^*$ represents the executed decisions that are obtained from:
    \begin{align}
        \bar p_t, \bar b_t, \bar e_t &= \argmax_{\substack{p_t, b_t, e_t  \\\in\mathcal{E}(e_{t-1}) }} \lambda_t (p_t - b_t) - C p_t + h^{\gamma_t}(\bar{Q}_t(e_t|\lambda_{t+1})) \\
        p^*_t, b^*_t, e^*_t &= \argmax_{\substack{p_t, b_t, e_t  \\\in\mathcal{E}(e_{t-1}) }} \lambda_t (p_t - b_t) - C p_t + h^{\gamma_t}(\hat{Q}_t(e_t|\lambda_t))
    \end{align}
    \end{subequations}
    where $h^{\gamma_t}(\cdot)$ represents the deployment of conformal decision strategy as given in~\eqref{eq:control_cc}.
\end{definition}

Accordingly, the step loss can be calculated as:
\begin{align}
    \ell_{t+1} = | \Delta V_{t}|.
    \label{eq:tloss2}
\end{align}

\begin{proposition}
    The temporal value error-based step loss $\ell_{t}$ defined in~\eqref{eq:tloss2} is monotonically non-decreasing in the conformal control variable~$\gamma$.
    \label{prop:tloss2}
\end{proposition}

Due to the space limit, we defer the proof of Proposition~\ref{prop:tloss2} to Appendix~\ref{app:tloss2}.
Proposition~\ref{prop:tloss2} indicates a direct relationship between the step loss $\ell_{t}$ and the  conformal control variable~$\gamma$, which is a desirable property for risk-aware decision-making.

Note that this loss is not directly determined by either the forecast value or the magnitude of forecast error, making it a more decision-aided calibration approach. However, this approach tends to amplify errors already present in the marginal opportunity value function since $\bar q_t$ is updated with $\lambda_{t+1}$ and $\hat q_{t+1}$, where~$\hat q_{t+1}$ itself contains forecast uncertainty.

\yw{The calibration method CC--prediction err depends on a heuristic mapping function~$f$, which performs effectively when the underlying estimates are reliable. In contrast, CC--value error is simpler to implement but generally suboptimal.
We compare the performance of these two methods in the case studies.}

\section{Case Study}\label{sec:Case}
\yq{
This section presents case studies evaluating the proposed conformal
risk-aware arbitrage strategy under realistic real-time market
conditions. We validate its risk management effectiveness by comparing it with representative baselines and examining performance under varying risk preferences. Extended sensitivity analyses, including broader parameter variations, are provided in Appendix~\ref{app:sensitivity} for completeness.}

\subsection{Setup}
\subsubsection{Data Description}
We evaluate arbitrage strategy performance using real-time market price data from the New York ISO (NYISO) for the New York City (NYC) zone during 2023~\cite{NYISO_price}. Value function forecaster training and uncertainty quantification are conducted using historical price data from 2021 to 2022. 
\yw{The price data is recorded at 5-minute intervals.}

\subsubsection{Parameter Settings and Metrics}
Our simulation features a storage unit with power rating~$P = 0.5 MW$ and capacity~$E = 1.0 MWh$ that starts and ends at 50\% SoC, with discharging and charging efficiency $\eta = 0.9$.
\yw{Regarding the value function forecaster~$\mathcal M$}, we evaluate algorithm performance under two forecasting scenarios using the model configuration from~\cite{baker2023transferable}: a good-accuracy forecaster ($R^2 = 0.4$) and a poor-accuracy forecaster ($R^2 = -0.4$). This setup enables assessment of the algorithm robustness across varying prediction quality levels \yw{as measured by $R^2$ coefficients}. 
For CC--prediction err, we construct the conservativeness calibration function in the form of $f:\hat{\gamma}_t \leftarrow \bar{\gamma}  \left(1 -\exp(-k  \Delta q)\right)$.
We use the cumulative profit as the performance metric: $\sum_{\tau=1}^{t}\lambda_{\tau}(p^*_{\tau}-b^*_{\tau}) - cp^*_{\tau}$, \yw{which measures the total profit gained over the time horizon $\{1,\ldots,t\}$}.

\subsection{Results}

\subsubsection{Baseline Algorithms}

We evaluate the two variants of our proposed conformal risk-aware algorithm against the baseline algorithms summarized in Table~\ref{tab:algorithms}, detailed in Appendix~\ref{app:baseline}. \yw{For benchmarking purposes, we also include the offline optimal profit, a non-causal upper bound computed in hindsight using the groundtruth prices and thus opportunity value functions.}
\yq{The detailed configuration details for all baseline algorithms are provided in the caption of Figure~\ref{fig:comp}.}

\subsubsection{Arbitrage Performance Comparison}

\begin{figure}[t]
    \centering
    \begin{subfigure}[b]{0.48\textwidth}
        \centering
        \includegraphics[width=\textwidth]{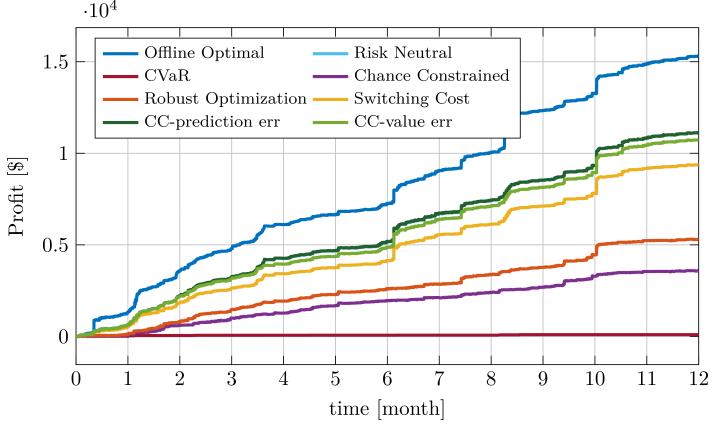}
        \caption{$R^2 = 0.4$.}
        \label{fig:comp_good}
    \end{subfigure}
  \hfill
    \begin{subfigure}[b]{0.48\textwidth}
        \centering
        \includegraphics[width=\textwidth]{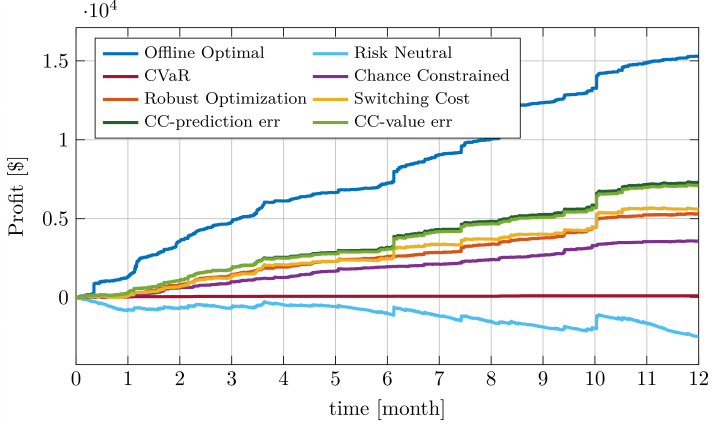}
        \caption{$R^2 = -0.4$.}
        \label{fig:comp_bad}
    \end{subfigure}
    \caption{Arbitrage performance comparison across different algorithms under varying opportunity value function forecast accuracy (measured by $R^2$): Offline Optimal represents the optimum assuming perfect knowledge; Risk Neutral deploys~\eqref{eq:arbitrage}; CVaR sets risk scaling factor~$\mu=1$; Chance Constrained assumes $\lambda_t$ following normal distribution and sets threshold~$\Gamma=0.6$; Robust Optimization estimates an ellipsoid uncertainty set and sets threshold~$\Gamma=1.0$; Switching Cost sets the penalty factor~$\zeta=400$; \yq{CC--prediction err sets the mapping function with~$\hat{\gamma} = 3.0$ and $k=0.1$; both CC--prediction err and CC--value err set~$\rho = 0.001$ and $\sigma = 10$.}}
    \label{fig:comp}
  \end{figure}

Figure~\ref{fig:comp} presents a comprehensive comparison of arbitrage performance across different algorithms under varying opportunity value function forecast accuracy levels: 
\yq{a reasonably good forecast ($R^2 = 0.4$) for general demonstration, and a poor forecast ($R^2 = -0.4$) for extreme stress testing.}

With good forecast accuracy ($R^2 = 0.4$) as shown in Figure~\ref{fig:comp}(\subref{fig:comp_good}), the proposed methods remain competitive with baseline approaches while providing additional risk protection. The risk-neutral strategy performs well under accurate predictions, as expected, but the conformal strategies maintain similar performance levels while offering downside protection. Notably, CC--prediction err closely tracks the risk-neutral strategy under good forecasting conditions, showing that the method adaptively reduces conservativeness when forecast accuracy improves. Overall, the conformal strategies demonstrate competitive performance under good forecasting conditions.

When forecast accuracy is poor ($R^2 = -0.4$) as shown in Figure~\ref{fig:comp}(\subref{fig:comp_bad}), both proposed conformal strategies (CC--prediction err and CC--value err) significantly outperform all baseline methods.
The conformal approaches recover approximately 47\% of the offline optimal profit while the risk-neutral baseline incurs substantial losses, demonstrating the effectiveness of \yq{conformal risk-aware decision-making} in managing uncertainty when predictions are unreliable. The conformal strategies exhibit superior performance under poor forecasting conditions.

  \begin{figure}[t]
    \centering
    \begin{subfigure}[b]{0.48\textwidth}
        \centering
        \vspace{.5mm}
        \includegraphics[width=\textwidth]{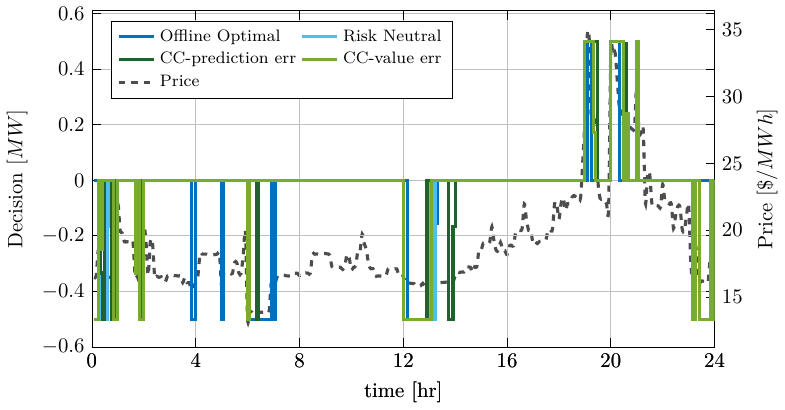}
        \caption{$R^2 = 0.4$.\vspace{-.5mm}}
        \label{fig:cc_good}
    \end{subfigure}
  \hfill
    \begin{subfigure}[b]{0.48\textwidth}
        \centering
        \includegraphics[width=\textwidth]{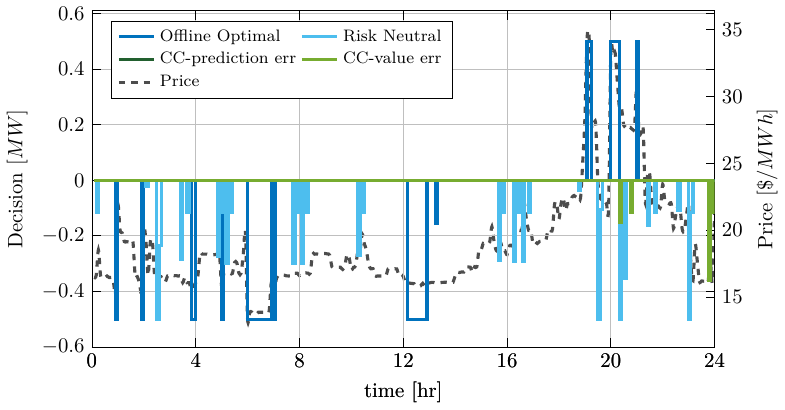}
        \caption{$R^2 = -0.4$.}
        \label{fig:cc_bad}
    \end{subfigure}
    \caption{Daily dispatch decisions comparing risk-neutral baseline and proposed conformal risk-aware strategies for good-accuracy ($R^2 = 0.4$) and poor-accuracy  ($R^2 = -0.4$) forecasters.}
    \label{fig:cc}
  \end{figure}
  
\subsubsection{Daily Decision Analysis}
Figure~\ref{fig:cc} provides detailed insight into daily dispatch decisions, comparing risk-neutral and conformal strategies under different forecast accuracy scenarios. 

Under reliable predictions as shown in Figure~\ref{fig:cc}(\subref{fig:cc_good}), CC--prediction err closely follows the offline optimal and risk-neutral strategies, particularly during high-value trading opportunities around hours 18-22 when prices peak above \$30/MWh. The strategy confidently captures discharge opportunities during price spikes while charging during low-price periods. CC--value err shows slightly more conservative behavior but maintains similar trading patterns.

The behavioral differences become pronounced under unreliable forecasting as shown in Figure~\ref{fig:cc}(\subref{fig:cc_bad}). The conformal strategies exhibit significantly more conservative decision-making with reduced trading frequency while the risk-neutral strategy continues aggressive trading based on poor forecasts. 
Both conformal methods show fewer trading actions, avoiding potentially harmful decisions based on poor forecasts.
For example, the risk-neutral strategy decides to costly mistakes such as charging during high-price periods around hours 18-22 when discharging would be optimal, the conformal strategies adopt a defensive posture.
This conservative behavior sacrifices some potential profits during uncertain periods but prevents the substantial losses that result from acting on misleading forecasts, ultimately leading to superior cumulative performance.

  \subsubsection{Impact of Risk Preference}
  \begin{figure}[t]
    \centering
    \begin{subfigure}[b]{0.49\textwidth}
        \centering
        \includegraphics[width=\textwidth]{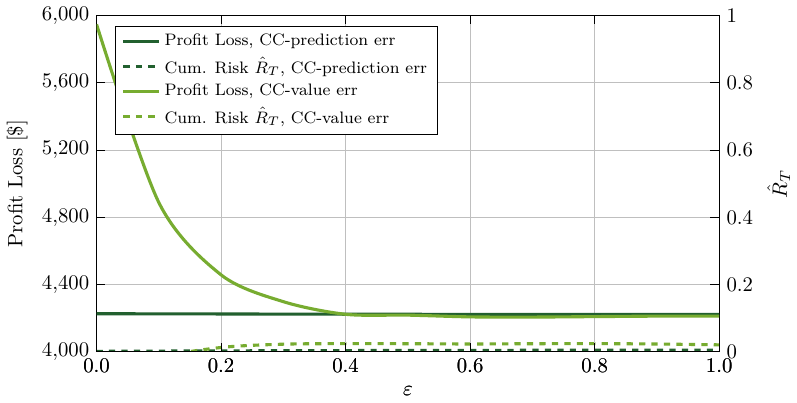}
        \caption{$R^2 = 0.4$.}
        \label{fig:risk_good}
    \end{subfigure}
  \hfill
    \begin{subfigure}[b]{0.49\textwidth}
        \centering
        \includegraphics[width=\textwidth]{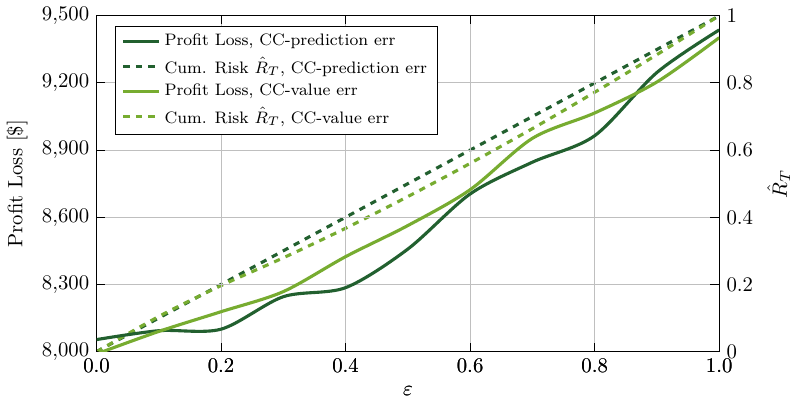}
        \caption{$R^2 = -0.4$.}
        \label{fig:risk_bad}
    \end{subfigure}
    \caption{Impact of risk tolerance $\varepsilon$ considering good-accuracy ($R^2 = 0.4$) and poor-accuracy  ($R^2 = -0.4$) forecasters.}
    \label{fig:risk}
  \end{figure}

  Figure~\ref{fig:risk} illustrates how the choice of risk tolerance $\varepsilon$ influences both realized profit loss (solid lines) and the cumulative risk $\hat{R}_T$ (dashed lines), as defined in Equation~\eqref{eq:regret} and Constraint~\eqref{eq:cum_risk}, under different forecast accuracy.

For the accurate forecaster case (Figure~\ref{fig:risk}(\subref{fig:risk_good})), profit loss remains relatively flat and the risk measure $\hat{R}_T$ stabilizes well below the risk threshold. 
This indicates that forecast errors are sufficiently small such that changes in risk tolerance $\varepsilon$ have little impact on cumulative risk.
However, we also observe that when $\varepsilon$ is set too conservatively (low risk tolerance), the storage misses profitable opportunities, resulting in a higher profit loss compared with more moderate settings.

In contrast, under poor forecast accuracy (Figure~\ref{fig:risk}(\subref{fig:risk_bad})), both profit loss and $\hat{R}_T$ increase almost linearly with $\varepsilon$. 
In this case, greater risk tolerance results in higher realized losses and a proportional rise in cumulative risk.
This linear relation between cumulative risk and risk tolerance $\varepsilon$ is consistent with the risk control target~\eqref{eq:cum_risk}, demonstrating that the conformal controller effectively regulates risk exposure.
Similarly, the results show a strong positive positive correlation between profit loss and $\hat{R}_T$, confirming that the ultimate profit loss and observable cumulative risk are closely correlated as Proposition~\ref{prop:bellman}.

It is also important to emphasize that the temporal loss definition~$\ell_t$, used to calculate cumulative risk, is calibration-method specific. Therefore, the relationship between profit loss and $\hat{R}_T$ is not guaranteed to be strictly monotonic across methods. For example, in Figure~\ref{fig:risk}(\subref{fig:risk_bad}), CC--value err yielding lower $\hat{R}_T$ does not necessarily lead to lower profit loss.

\yq{
\section{Discussion and Future Work}\label{sec:Limitations}

\subsection{Linear Degradation Modeling}
Our framework currently models battery degradation as a constant marginal operational cost $C$. This linear approximation preserves the convexity of the optimization problem, enabling the derivation of the computationally efficient analytical dispatch policies in Eq.~\eqref{eq:control_cc}. While this simplification is standard in literature~\cite{xu2020,baker2023transferable,yi2025,wu2025,yi2025decision}, we recognize that realistic battery degradation is often nonlinear, exhibiting increased marginal costs at extreme states of charge or higher discharge rates. 
Adopting a nonlinear model would alter the marginal opportunity value function, leading to smoother continuous dispatch responses rather than the threshold-based behavior observed under linear costs.

However, the overall robustness of the proposed conformal framework remains intact.
First, the theoretical guarantees for risk control (Theorem~\ref{theorem:cc}) and the validity of the temporal difference error as a proxy (Proposition~\ref{prop:bellman}) rely on the boundedness and suitable monotonicity of the calibration loss, not on the specific form of the operational cost function. Second, the core idea of our conformal risk-aware framework remains applicable, as its role is to adjust decision conservativeness in response to forecast quality. That said, omitting nonlinear degradation may introduce systematic bias in the value function that the controller must counteract by widening prediction sets.

Future work will extend the conformal risk controller to accommodate these nonlinear or numerical optimization settings, thereby allowing the framework to simultaneously manage market price uncertainty and complex physical aging processes.

\subsection{Symmetric Uncertainty Assumptions}
    For simplicity, the prediction set $\hat{C}_t$~\eqref{eq:cc_set} is constructed using symmetric quantile bounds, which assume approximately symmetric forecast errors. This design choice facilitates implementation and analytical tractability. However, given that electricity prices often exhibit skewed or log-normal distributions, the conformal risk management framework can be readily extended to more informative asymmetric prediction sets by selecting distinct upper and lower quantile levels according to specific risk preferences, as illustrated for example in~\cite{pouplin2025}. We highlight this as an important direction for future refinement of the controller design.

}

\section{Conclusions}\label{sec:Conclusion}
This paper presents a conformal risk-aware framework for online energy storage arbitrage under prediction uncertainty. We establish temporal difference error as a proxy for unobservable profit losses and develop two online calibration strategies based on prediction error and value error that adjust conformal control variables. The proposed conformal controller achieves bounded cumulative risk with convergence guarantees under minimal distributional assumptions while maintaining computational efficiency for real-time implementation. Extensive empirical validation demonstrates robust performance across varying forecast accuracy levels through adaptive conservativeness calibration. The framework's distribution-free uncertainty quantification, online adaptation capabilities, and statistical coverage guarantees enable deployment in real-world energy storage operations and extend to broader sequential decision-making problems under imperfect predictions.

\section*{Acknowledgements}
JA acknowledges funding from the NSF grants ECCS 2144634 and 2231350 and the Columbia Data Science Institute. BX acknowledges funding from the NSF grant ECCS 2239046.

\balance
\bibliographystyle{ACM-Reference-Format}
\bibliography{uncertainty}

\appendix
\yw{
\section{Conformal Prediction}\label{app:conformal_prediction}
Conformal prediction~\cite{angelopoulos2022} is a  distribution-free uncertainty quantification paradigm. In contrast to methods reliant on specific distributional or model assumptions, conformal prediction provides explicit, non-asymptotic statistical guarantees applicable to any \emph{pretrained prediction model} $\hat{f}$, such as a neural network.

The most widely-used variant, split-conformal prediction, constructs a \emph{prediction set}, $\hat{C}(X_{\mathrm{test}})$, for a new test input $X_{\mathrm{test}}$. This set is guaranteed to contain the true, unknown label $Y_{\mathrm{test}}$ with a \emph{user-specified probability}. This property, termed marginal coverage, is formally expressed as:
$$\mathbb{P}(Y_{\mathrm{test}} \in \hat{C}(X_{\mathrm{test}})) \ge 1-\varepsilon.$$
Here, $\varepsilon \in [0, 1]$ is the user-defined error tolerance. This guarantee holds under the sole assumption that the calibration and test data are exchangeable.

The split-conformal procedure is implemented as follows:
\begin{enumerate}
    \item Data Partition: A portion of the available data is held out to form a \emph{calibration set}, $\{(X_i, Y_i)\}_{i=1}^n$, which must not have been used during model training.
    
    \item Score Function Definition: A \emph{nonconformity score function}, $s(x, y) \in \mathbb{R}$, is selected for any input $x$ and output $y$. This function quantifies the degree of disagreement between an input $x$ and a candidate label $y$, based on the model's output $\hat{f}(x)$. Larger scores indicate higher model error or uncertainty.
    
    \item Score Computation: The score function is applied to all data points in the calibration set, yielding $n$ calibration scores: $s_i = s(X_i, Y_i)$ for $i=1, ..., n$.
    
    \item Quantile Determination: A threshold, $\hat{k}$, is determined by calculating the $\frac{\lceil(n+1)(1-\varepsilon)\rceil}{n}$ empirical quantile of calibration scores $\{s_i\}_{i=1}^n$, where $\lceil \cdot \rceil$ is the ceiling function.
    
    \item Prediction Set Construction: For a new test input $X_{\mathrm{test}}$, the prediction set $\hat{C}(X_{\mathrm{test}})$ is constructed by including all possible labels $y$ for which the nonconformity score does not exceed the threshold $\hat{k}$:
    $$\hat{C}(X_{\mathrm{test}}) = \{y : s(X_{\mathrm{test}}, y) \le \hat{k}\}.$$
\end{enumerate}

The coverage guarantee is valid for any choice of model or score function, detailed proof available in \cite[Appendix~D]{angelopoulos2022}. The selection of an appropriate score function, however, is critical as it dictates the efficiency and utility (e.g., the average size or adaptivity) of the resulting prediction sets. In practice, typical choices include absolute residuals in regression tasks or one minus the softmax probability of the true class in classification tasks.

}
\section{Analytical Solution to Energy Storage Arbitrage}\label{app:arbitrage_solution}
The following results from~\cite{zheng2022} are reproduced here for reader completeness and to facilitate interpretation of our risk-aware dispatch policy~\eqref{eq:control_cc}.

The optimal solutions~$(p^*_t,b^*_t)$ for all $t \in \mathcal{T}$ to~\eqref{eq:arbitrage} can be obtained using the following dispatch policy given the realized price~$\lambda_t$, the prediction of marginal opportunity value function~$\hat{q}_t(e_t|\lambda_t)$ and current SoC~$e_{t-1}$:
\begin{subequations}
\begin{align}
  p^*_t & = \min\{\hat{p}_t, e_{t-1} \eta\}\label{eq:decision_cap}\\
  b^*_t & = \min\{\hat{b}_t, (E - e_{t-1})/\eta\} \label{eq:decision_cap_b}
\end{align}
where \( \hat{p}_t \) and \( \hat{b}_t \) are calculated based on price thresholds:
\begin{align}
(\hat{p}_t, \hat{b}_t) =
\begin{cases} 
(0, P) & \text{if } \lambda_t \leq \hat{q}_t(e_{t-1} + P \eta) \eta \\
(0, \alpha) & \text{if } \hat{q}_t(e_{t-1} + P \eta) \eta < \lambda_t \leq \hat{q}_t(e_{t-1}) \eta \\
(0, 0) & \text{if } \hat{q}_t(e_{t-1}) \eta < \lambda_t \leq [\hat{q}_t(e_{t-1})/\eta + C]^+ \\
(\beta, 0) & \text{if } [\hat{q}_t(e_{t-1})/\eta + C]^+ < \lambda_t \\
&\qquad  \leq [\hat{q}_t(e_{t-1} - P/\eta)/\eta + C]^+ \\
(P, 0) & \text{if } \lambda_t > [\hat{q}_t(e_{t-1} - P/\eta)/\eta + C]^+
\end{cases}
\label{eq:control0}
\end{align}
where $[\cdot]^+ = \max \{ \cdot, 0 \}$ indicates positive projection, and~\( \alpha \) and \( \beta \) are given by:
\begin{align*}
\alpha = \frac{\hat{q}_t^{-1}\left(\lambda_t/\eta\right) - e_{t-1}}{\eta},\quad \beta = \left(e_{t-1} - \hat{q}_t^{-1}\left(\left(\lambda_t - C\right) \eta\right)\right){\eta} 
\end{align*}
\label{eq:control}
\end{subequations}
where \( \hat{q}_t^{-1}(\cdot) \) is the inverse function of \( \hat{q}_t (\cdot)\). 

The dispatch policy~\eqref{eq:control} reads as follows. Equations~\eqref{eq:decision_cap} and~\eqref{eq:decision_cap_b} enforce capacity constraints over the discharge \( \hat{p}_t \) and charge \( \hat{b}_t \) decisions. Equation~\eqref{eq:control0} calculates control decisions by balancing the trade-off between the immediate payoff and the change in future opportunity. In the first case, the unit charges at full power when the gain in future opportunity exceeds the cost of charging accounting for the efficiency loss. The fifth case follows similar logic. In the second case, the unit charges at partial power such that the resulting marginal opportunity value equals the marginal objective value. This establishes an equilibrium point through the inverse marginal value function. The fourth case follows similar logic. In the third case, the unit remains idle when neither charging nor discharging provides sufficient value to compensate for the associated costs and efficiency losses.

\yq{
\section{Proof of Proposition~\ref{prop:bellman}} \label{app:bellman}
}
\begin{proof}
Under Assumption~\ref{assump:unique}, the arbitrage problem~\eqref{eq:arbitrage} has unique optimal solution.
That's to say, the transition from current state~$e_t$ to action~$(p_{t+1}, b_{t+1})$ and next state~$e_{t+1}$ is deterministic. By definition $\Delta(e_t)~=~\hat{Q}_t(e_t|\lambda_t)~-~Q_t(e_t)$, we can decompose the value error as: 
\begin{align*}
    \Delta(e_t) &= \hat{Q}_t(e_t|\lambda_{t}) - Q_t(e_t) \\
    &= \hat{Q}_t(e_t|\lambda_{t}) - (\Pi_t + Q_t(e_{t+1})) \\
    &= \hat{Q}_t(e_t|\lambda_{t}) - (\Pi_t + \hat{Q}_{t+1}(e_{t+1}|\lambda_{t+1})~-~\Delta(e_{t+1}))\\
    &= \hat{Q}_t(e_t|\lambda_{t}) - (\Pi_t + \hat{Q}_{t+1}(e_{t+1}|\lambda_{t+1}))+~\Delta(e_{t+1})\\
    &= \delta(e_{t})+~\Delta(e_{t+1}) \qquad \text{(by definition of $\delta(e_t)$)}
\end{align*}

$(\Rightarrow)$ First we show that if $\delta(e_t) = 0$, then $\Delta(e_t) = 0$, for all $e_t \in [ 0,E], \ t \in \mathcal T$. 
\yq{Given the recursive relationship established above, for a given time step $t$, $\Delta(e_t) = \delta(e_{t}) + \Delta(e_{t+1}) = 0$ if $\Delta(e_{t+1}) = 0$ for all $e_{t+1} \in [0,E]$, since $\delta(e_{t}) = 0$.
With the boundary condition $\Delta(e_T) = 0$ for all $e_T \in [0,E]$ at $t = T$, 
it follows by induction that $\Delta(e_t) = 0$ if $\delta(e_t) = 0$ for all $e_t \in [0,E], \ t \in \mathcal T$.}

$(\Leftarrow)$ Next, we show that if $\Delta(e_t) = 0$, then $\delta(e_t) = 0$, for all $e_t \in [ 0,E], \ t \in \mathcal T$. Similarly, we can have $\Delta(e_{t}) = 0 \implies 0 = \delta(e_{t}) + 0 \implies \delta(e_{t}) = 0$.
\end{proof}

\section{Proof of Theorem~\ref{theorem:cc}} \label{app:theorem_cc}

\begin{proof}

By the definition of the update rule~\eqref{eq:cc}, we have 
\[
\gamma_{t+1} = \gamma_1 + \rho \sum_{s=1}^{t} (\varepsilon - \ell_s),
\]
where $\ell_s$ is the arbitrage loss of interest (normalized to $[0,1]$) incurred by the dispatch decision at time step $s$.  

Rearranging the update rule gives
\begin{align}
\hat{R}_t(\gamma_{1:t}) 
= \frac{1}{t} \sum_{s=1}^{t} \ell_s 
= \varepsilon + \frac{\gamma_1 - \gamma_{t+1}}{\rho t}. 
\label{eq:risk_bound_cc1}
\end{align}

Next, we establish a lower bound on $\gamma_t$ for all $t \in \mathcal T^+$.  
Intuitively, $\gamma_t$ regulates the aggressiveness of dispatch decisions, and bounding it away from $-\infty$ ensures that the storage dispatch remains economically \emph{eventually safe}.

Assume by contradiction that there exists $\inf_{s \in \mathcal T^+} \gamma_s < \gamma^{\text{safe}} - K\rho$. 
Let $t\in \mathcal T^+$ be the first time step such that $\gamma_t$ falls below this threshold, i.e.,
\[
\gamma_t < \gamma^{\text{safe}} - K\rho
\quad \text{and} \quad 
\gamma_s \geq \gamma^{\text{safe}} - K\rho \;\; \forall s < t,\, s \in \mathcal T^+.
\]
This implies $\gamma_t < \gamma^{\text{safe}} - K\rho \leq \gamma_s$ for all $s < t$.

We now show that $t > K$ must hold.  
Assume by contradiction that $t \leq K$. 
By the update rule~\eqref{eq:cc}, the maximal sequential change in $\gamma_s$ is bounded as $\sup_{s \in \mathcal T} |\gamma_{s+1} - \gamma_s| < \rho$ given normalized $\ell_s \in [0,1]$.

First, we prove that~\eqref{eq:risk_bound_cc} holds if $\gamma_1 \geq \gamma^{\text{safe}} - \rho$.

We can recursively apply the update rule~\eqref{eq:cc} and obtain:
\[
\gamma_1 < \gamma^{\text{safe}} - \rho - (K-t)\rho < \gamma^{\text{safe}} - \rho,
\]
which contradicts the assumption that $\gamma_1 \geq \gamma^{\text{safe}} - \rho$.  
Therefore, $t > K$.

Since $t > K$, we can derive:
\begin{align}
\gamma_{t-k} < \gamma^{\text{safe}} - (K-k)\rho \leq \gamma^{\text{safe}}, 
\quad \forall k \in \{0, \ldots, K\}. 
\label{eq:gamma_t}
\end{align}

By Definition~\ref{def:1} (eventually safe dispatch), this implies that the arbitrage cumulative risk across the preceding $K$ dispatch time steps satisfies
\begin{align}
\frac{1}{K} \sum_{k=1}^{K} \ell_{t-k} \leq \varepsilon^{\text{safe}}.
\label{eq:ell_t_k}
\end{align}

By recursively applying the update rule, and combining~\eqref{eq:gamma_t} and~\eqref{eq:ell_t_k}, we have 
\begin{align*}
    \gamma_t &= \gamma_{t-K} + K\rho \left( \varepsilon - \frac{1}{K} \sum_{k=1}^{K} \ell_{t-k} \right) \\
    &\geq \gamma_{t-K} + K\rho(\varepsilon - \varepsilon^{\text{safe}}) \\
    &\geq \gamma_{t-K}.
\end{align*}

Since $t$ is the first time step where $\gamma_t < \gamma^{\text{safe}} - K\rho$, and we have shown that $\gamma_t \geq \gamma_{t-K}$, we arrive at a contradiction.

Therefore, $\gamma_t \geq \gamma^{\text{safe}} - K\rho$ for all $t \in \mathcal T^+$.  
Substituting this bound into~\eqref{eq:risk_bound_cc1} yields
\begin{align*}
\hat{R}_t(\gamma_{1:t}) 
\;\leq\; \varepsilon + \frac{(\gamma_1 - \gamma^{\text{safe}})/\rho + K}{t},
\end{align*}
which satisfies~\eqref{eq:risk_bound_cc} and certifies that the arbitrage cumulative risk of the energy storage dispatch converges within the prescribed tolerance under eventual safety.  

Next, we relax the assumption on $\gamma_1$ and prove that~\eqref{eq:risk_bound_cc} still holds.
We establish the result by examining two scenarios. 

\text{Scenario 1: all early $K$ steps are safe.}   Assume $\gamma_k \leq \gamma^{\text{safe}}$ for all $k$ in  $K$ steps.  
Since the dispatch is eventually safe during these periods, each per-step loss of interest $\ell_t$ is bounded by the safe tolerance $\varepsilon^{\text{safe}}$. Hence
\begin{align}
   \frac{1}{T}\sum_{t=1}^{K}\ell_t \;\leq\; \frac{K\varepsilon^{\text{safe}}}{T}. \label{eq: case1}
\end{align}

Since the definition of eventual safety is satisfied within the first 
$K$ steps, we can apply \eqref{eq:risk_bound_cc} from step 
$K+1$ onward. For every $t \in \{K,\dots,T\}$,
\begin{align}
\hat{R}_t(\gamma_{1:t}) \;\leq\; \varepsilon \;+\; \frac{(\gamma_1 - \gamma^{\text{safe}})/\rho + K}{t}.
\end{align}

Thus the cumulative risk satisfies
\begin{align}
   \frac{1}{T}\sum_{t=K+1}^{T}\ell_t \;\leq\; \varepsilon + O(1/T).  \label{eq: case2}
\end{align}

Combining the \eqref{eq: case1} and \eqref{eq: case2}, we obtain
\[
\hat{R}_T
= \frac{1}{T}\sum_{t=1}^{K}\ell_t
  + \frac{1}{T}\sum_{t=K+1}^{T}\ell_t
\;\leq\; \varepsilon + \frac{K(\varepsilon^{\text{safe}}-\varepsilon)}{T} + O(1/T).
\]
Therefore, as $T \to \infty$, the bound
$
\hat{R}_T \;\leq\; \varepsilon + O(1/T)
$
holds.

\medskip

Scenario 2: an unsafe step occurs early at step $k$.
Suppose there exists some $k \le K$ such that $\gamma_k > \gamma^{\text{safe}}$, and let $k$ be the first such index. 
For $t=1,\dots,k-1$, since the loss is normalized to $[0,1]$, we can bound
\begin{align}
    \frac{1}{T}\sum_{t=1}^{k-1}\ell_t \;\le\; \frac{k-1}{T}. \label{eq:case3}
\end{align}

From $t=k$ onward, we apply the risk bound~\eqref{eq:risk_bound_cc} for all $t \in \{k,\dots,T\}$,
\begin{align}
  \hat{R}_t(\gamma_{1:t}) 
\;\le\;
\varepsilon 
\;+\; 
\frac{(\gamma_k - \gamma^{\text{safe}})/\rho + (T-k+1)}{t}.  \label{eq:case4}
\end{align}

Here, the term $(T-k+1)$ provides a loose upper bound before safety is reestablished, since at worst every step from $k$ to $T$ could incur the maximum loss.

Therefore, the cumulative risk from $t=k$ onward is bounded by the expression above, while $t<k$ contributes at most $(k-1)/T$. Combining \eqref{eq:case3} and \eqref{eq:case4} yields
\[
\hat{R}_T(\gamma_{1:T}) 
\;\le\;
\varepsilon 
\;+\;
\frac{k-1}{T}
\;+\;
\frac{(\gamma_k - \gamma^{\text{safe}})/\rho + (T-k+1)}{T}.
\]
Since $k$ is finite and the additional terms vanish as $T \to \infty$, we get
$
\hat{R}_T(\gamma_{1:T}) \;\le\; \varepsilon + O(1/T)$.

Consider both scenarios, we conclude the proof.

\qedhere
 \end{proof}

 \section{Proof of Proposition~\ref{prop:optimal_width}} \label{app:optimal_width}

\begin{proof}
    We start with a simplified case using 1) single-segment marginal value functions for each time step $t$, i.e., $ q_t = q_t(e_t)|_{e_t\in [0,E]}$; 2) perfect efficiency $\eta = 1$.

    Accordingly, the dispatch policy~\eqref{eq:control} will be modified to:
    {\small\begin{align*}
        (p^*_t,b^*_t)^{\gamma} =  \begin{cases} 
        \left(0, \min\{P, E - e_{t-1}\}\right) & \text{if } \lambda_t \leq \min\{\hat{C}_t\}  \\
        \left( \min\{P, e_{t-1}\}, 0\right) & \text{if } \lambda_t > [\max\{\hat{C}_t\} + C]^+ \\
        (0, 0) & o.w.
        \end{cases}
      \end{align*}}

    According to
    \yq{Section~\ref{sec:problem_statement}},
     a dispatch policy that can maximize the total profit maintains an optimal total value constantly at any time step $t$, i.e., $V_t:= \Pi_t(p'_t,b'_t|\lambda_t)  + Q_t(e'_{t+1})$, where $\Pi_t$ and $e'_{t+1}$ are updated with: 
    {\small\begin{align*}
        (p'_t,b'_t) =   \begin{cases} 
        \left(0, \min\{P, E - e_{t-1}\}\right) & \text{if } \lambda_t \leq q_t  \\
        \left( \min\{P, e_{t-1}\}, 0\right) & \text{if } \lambda_t > [q_t + C]^+ \\
        (0, 0) & o.w.
        \end{cases}
      \end{align*}}
      and the suboptimal value is given by $\bar{V}_t:= \Pi_t(p^*_t,b^*_t|\lambda_t)  + Q_t(\bar{e}_{t+1})$.

Therefore, the profit loss incurred is given by:
\begin{align}
    \Delta V_t = & V_t - \bar{V}_t  \nonumber \\
    =  &\lambda_t \left((p'_t - b'_t) - (p^*_t - b^*_t) \right) - C (p'_t - p^*_t) \nonumber\\ 
    &+ Q_t(e'_{t+1}) - Q_t(\bar{e}_{t+1}) \qquad \text{(by definition)}\nonumber\\
    =  &\lambda_t \left((p'_t - b'_t) - (p^*_t - b^*_t) \right) - C (p'_t - p^*_t) \nonumber\\ 
    &+ q_t(e'_{t+1}-\bar{e}_{t+1}) \quad \text{(single-segment $q_t$ assumption)}\nonumber\\
    =  &\lambda_t \left((p'_t - b'_t) - (p^*_t - b^*_t) \right) - C (p'_t - p^*_t) \nonumber\\ 
    &+ q_t\left(-(p'_t - b'_t) + (p^*_t - b^*_t) \right) \qquad \text{(SoC evolution)}\nonumber\\
    =  &(\lambda_t- C - q_t) (p'_t - p^*_t) - (\lambda_t -q_t) (b'_t - b^*_t)  
    \label{eq:vloss}
\end{align}

There are four cases with respect to the value of $\Delta V_t$:
\begin{enumerate}
    \item \textbf{False Buy (Type I Loss)}
    When $\lambda_t \in (q_t, \min\{\hat{C}_t\}]$, the conformal policy buys but optimal policy holds/sells.
    \item \textbf{False Sell (Type I Loss)}
    When $\lambda_t \in ([q_t + C]^+, [\max\{\hat{C}_t\} + C]^+]$, the conformal policy sells but optimal policy holds/buys.
    \item \textbf{Missed Buy (Type II Loss)}
    When $\lambda_t \in [\min\{\hat{C}_t\}, q_t)$, the conformal policy holds but optimal policy buys.
    \item \textbf{Missed Sell (Type II Loss)}
    When $\lambda_t \in [\max\{\hat{C}_t\} + C]^+, [q_t + C]^+)$, the conformal policy holds but optimal policy sells.
\end{enumerate}

The expected profit loss is:
\begin{align}
\mathbb{E}[\Delta V_t] = \mathbb{E}[\text{Type I Loss}] + \mathbb{E}[\text{Type II Loss}] \label{eq:Vloss}
\end{align}
where $\mathbb{E}[\text{Type I Loss}]$ and $\mathbb{E}[\text{Type II Loss}]$ correlate to  False Actions and Missed Actions, respectively, and can be calculated based on~\eqref{eq:vloss}.

Consider the extreme cases: 
\begin{itemize}
  \item When $\gamma_t \to 0^+$,  the prediction set widens, and $\mathbb{E}[\Delta V_t]$ reduces to the total of $\mathbb{E}[\text{Type II Loss}]$, which is  finite and strictly positive value.
  \item When $\gamma_t \to +\infty$, the prediction set narrows, and $\mathbb{E}[\Delta V_t]$ is dominated by $\mathbb{E}[\text{Type I Loss}]$ and is strictly positive.
\end{itemize}

If there exists an optimal $\gamma^*_t$, then  $\gamma^*_t$ satisfies the first-order condition $\mathrm{d} \mathbb{E}[\Delta V_t] / \mathrm{d}\gamma_t = 0$, this leads to: 
\[
\frac{\mathrm{d}}{\mathrm{d}\gamma_t}\mathbb{E}[\text{Type I Loss}] + \frac{\mathrm{d}}{\mathrm{d}\gamma_t}\mathbb{E}[\text{Type II Loss}] = 0.
\]

Consider that when $\gamma_t$ increases, the prediction set becomes narrower, leading to:
\begin{itemize}
    \item Type I Loss increases with more false actions
    \item Type II Loss decreases with fewer missed actions
\end{itemize}

Combining the two extreme cases above, we have dominant $\mathrm{d}\mathbb{E}[\text{Type I Loss}]/\mathrm{d}\gamma_t$ as $\gamma_t \to +\infty$ and dominant $\mathrm{d}\mathbb{E}[\text{Type II Loss}]/\mathrm{d}\gamma_t$ as $\gamma_t \to 0^+$. Given the continuity of $\mathbb{E}[\Delta V_t]$ in $\gamma_t$ as implied by~\eqref{eq:vloss}, there exists at least one optimal $\gamma^*_t$ such that $\mathrm{d} \mathbb{E}[\Delta V_t] / \mathrm{d}\gamma_t = 0$.

Next, we assess the sensitivity of the optimal parameter $\gamma_t^*$ to the uncertainty parameter $\epsilon_t$ by:
\[
\frac{d\gamma_t^*}{d\epsilon_t} = - \frac{\partial^2 \mathbb{E}[\Delta V_t]
 / \partial\gamma_t \partial\epsilon_t}{\partial^2 \mathbb{E}[\Delta V_t]
 / \partial\gamma_t^2}
\]
Given $\frac{\partial^2 \mathbb{E}[\Delta V_t]}{\partial \gamma_t^2} \ge 0$, the sign of this derivative is determined by $\frac{\partial^2 \mathbb{E}[\Delta V_t]}{\partial\gamma_t \partial\epsilon_t}$.

As discussed above, $\frac{\partial \mathbb{E}[\Delta V_t]}{\partial \gamma_t} \rightarrow 0$ as $\gamma_t \rightarrow \gamma^*_t$, which means that these effects are balanced.
For $\frac{\partial \mathbb{E}[\Delta V_t]}{\partial \epsilon_t}$, this correlation is naturally positive, as larger forecast errors in turn increase the expected profit loss.
Therefore we have approximately $\frac{\partial^2 \mathbb{E}[\Delta V_t]}{\partial\gamma_t \partial\epsilon_t} \ge 0$.

Thus, according to the implicit function theorem, we have an approximate expression for the sensitivity of the optimal parameter:
$
\frac{d\gamma_t^*}{d\epsilon_t} = - \frac{(+)}{(+)} \le 0
$.
The optimal control parameter $\gamma_t^*$ is non-increasing in forecast uncertainty parameter $\epsilon_t$ in expectation.

    The discussion above can be similarly extended to relaxed scenarios with multi-segment marginal value functions, various forecast error patterns and imperfect efficiency. 
\end{proof}

\section{Proof of Proposition~\ref{prop:tloss2}} \label{app:tloss2}

\begin{proof}
    The definition of conformal prediction set~\eqref{eq:cc_set} suggests that a decrease in $\gamma_t$ leads to a wider prediction set.
For the extreme case, as $\gamma_t \to 0^+$: $\lim_{\gamma_t \to 0^+} | \Delta V_{t}| = 0$, suggesting that small $\gamma_t$ makes the conformal strategy converge to the idle action $(0,0)$ regardless of the underlying value function, which provides a loose monotonicity in the step loss $\ell_{t}$. 
    
If we consider fixed environment, e.g., $\lambda_{t+1}$, we can express~\eqref{eq:Vloss2} as $\Delta V_t:=\Delta V_t(\gamma, \hat{Q})$, where we neglect the time index for simplicity. Similarly, we can express the decision-making policy as $d := (p_t, b_t) = \mathcal{D}_t(\gamma, Q)$, where $Q$ can take $\hat Q$, $\bar Q$ or other input given respective context.

\yq{Following Proposition~\ref{prop:optimal_width}},
we can state that a conservative policy is less sensitive to changes in forecast uncertainty, and vice versa. That is, for any two value functions $Q_1, Q_2$ and $0<\gamma_1 < \gamma_2$, the following holds:
\begin{align*}
    |\mathcal{D}_t(\gamma_1, Q_1) - \mathcal{D}_t(\gamma_1, Q_2)| \leq |\mathcal{D}_t(\gamma_2, Q_1) - \mathcal{D}_t(\gamma_2, Q_2)|
\end{align*}

The step loss $\Delta V_t > 0 $ only when there is a difference in the dispatch decisions made by the two strategies. A forecast update that is significant enough to change the decisions for $\gamma_1$ must also be significant enough to change that for $\gamma_2$

Notice the linear structure of the objective function in decisions $(p_t, b_t)$ and bounded constraints, we can assume there exists a constant $k$ such that
$
|\Delta V_t(\gamma)| = k |\mathcal{D}_t(\gamma, \bar{Q}_t) - \mathcal{D}_t(\gamma, \hat{Q}_t)|
$.
Then we have:
\begin{align*}
|\Delta V_t(\gamma_1)| &= k \cdot |\mathcal{D}_t(\gamma_1, \bar{Q}_t) - \mathcal{D}_t(\gamma_1, \hat{Q}_t)| \quad \text{(by linearity)} \\
&\le k \cdot |\mathcal{D}_t(\gamma_2, \bar{Q}_t) - \mathcal{D}_t(\gamma_2, \hat{Q}_t)| \\
&= |\Delta V_t(\gamma_2)| \quad \text{(by linearity)}
\end{align*}

Therefore, $\ell_{t}$ defined in~\eqref{eq:tloss2} is monotonically non-decreasing in~$\gamma$.

\end{proof}

\section{Baseline Algorithms}\label{app:baseline}
We compare our proposed algorithm with the following baseline algorithms:
\begin{enumerate}
    \item Risk-Neutral: decision-making according to the given value function forecast as defined in \eqref{eq:arbitrage}, 
    \item CVaR: traditional CVaR-based risk management approaches that minimize expected losses in the worst-case scenarios to provide downside protection, but depend on accurate scenario generation and cannot adapt online to changing market conditions. Let $\omega \in \Omega$ be the scenarios and $\varrho_{\omega}$ be the probability of each scenario, $\nu \in (0,1) $ be the confidence level, $\mu \in [0,1]$ be the risk scaling factor, and $\varphi_\omega$ be a nonnegative auxiliary variable:
      \begin{align*}
            \maximize{\substack{p_t, b_t, e_t  \\ \in\mathcal{E}(e_{t-1}) }} &\quad  \sum_{\omega\in \Omega} \varrho_{\omega}V_{t,\omega}  + \mu \text{CVaR}_{\nu} \\
    \text{s.t.} & \quad  V_{t,\omega} = \lambda_t (p_{t,\omega} - b_{t,\omega}) - C p_{t,\omega} + \hat Q_{t,\omega}(e_{t,\omega}|\lambda_t) \\
            & \quad \text{CVaR}_{\nu} = \text{VaR} - \frac{1}{1-\nu} \sum_{\omega \in \Omega} \varrho_\omega \varphi_\omega \\
            & \quad 0 \leq \varphi_\omega \geq \text{VaR} - V_{t,\omega} 
         \end{align*}

     \item Chance Constrained: chance-constrained optimization ensuring constraint violations occur with probability below specified thresholds~$\Gamma$, but requiring accurate probability distributions and lacks adaptability to distributional shifts~\cite{wu2025}. Let $\xi$ be an epigraph auxiliary variable:
      \begin{align*}
        \maximize{\substack{p_t, b_t, e_t  \in\mathcal{E}(e_{t-1}),\xi }}  \quad  &\xi \\
           \mathrm{s.t.} \quad & \mathbb{P}( \textstyle\sum_{t \in \mathcal{T}}\lambda_t (p_t-b_t)- Cp_t \geq \xi) \ge \Gamma 
         \end{align*}
    \item Robust Optimization: feasible solutions guarantee under worst-case realizations within predefined uncertainty sets~$\mathcal{U}_t$, but tends conservative and cannot adapt uncertainty set specifications online~\cite{wu2025}:
        \begin{align*}
        \maximize{\substack{p_t, b_t, e_t  \in\mathcal{E}(e_{t-1}), \xi }}  \  &\xi \\
        \mathrm{s.t.} \ & \underset{\lambda_t \in \mathcal{U}_t}{\text{minimize}}  \ \textstyle\sum_{t \in \mathcal{T}}\lambda_t (p_t-b_t) - Cp_t \geq \xi
      \end{align*}
   
    \item Switching Cost: an online prediction-based algorithm conversion with switching costs to handle noisy predictions~\cite{chen2015}, though its theoretical risk management guarantees assume observable prediction errors, which are not available in our setting. Let $\zeta$ be the switching cost penalty factor:
      \begin{align*}
        \maximize{\substack{p_t, b_t, e_t  \in\mathcal{E}(e_{t-1}) }} \quad \lambda_t (p_t - b_t) - C p_t + \hat Q_t(e_t|\lambda_t) - \zeta \|e_t - e_{t-1}\|
      \end{align*}

\end{enumerate}

\yq{
  \section{Conformal Controller Sensitivity Analysis}\label{app:sensitivity}
  We include here an extended sensitivity analysis of key algorithm parameters, including the controller learning rate $\rho$ in Eq.~\eqref{eq:cc} (Figure~\ref{fig:rho}) and the sensitivity parameter $\sigma$ in Eq.~\eqref{eq:cc_set} (Figure~\ref{fig:sigma}). The layout of each figure is similar to that of Figure~\ref{fig:risk}.

\subsection{Impact of Controller Learning Rate}

We analyze the sensitivity of the proposed algorithm to the conformal controller calibration learning-rate parameter $\rho$ in Eq.~\eqref{eq:cc}, which governs how quickly the control variable adjusts in response to observed losses.

Under good forecast quality (Figure~\ref{fig:rho}(\subref{fig:rho_good})), both CC-prediction err and CC-value err exhibit relatively mild variation across the tested range of $\rho$. This robustness arises because accurate forecasts stabilize the temporal-difference signal that drives calibration updates. Note that CC-value err shows slightly higher sensitivity to $\rho$ than CC-prediction err, as the the prediction-error-based updates leverage the heuristic conservativeness calibration function $f$, which helps guide the controller toward more stable convergence.

Under poor forecast quality (Figure~\ref{fig:rho}(\subref{fig:rho_bad})), larger learning rates cause the controller to update too aggressively. As $\rho$ increases, both the cumulative risk $\hat R_T$ and realized profit loss increase significantly for both the prediction-error-based and value-error-based calibration methods. This indicates that when forecasts are noisy, overly large $\rho$ magnifies volatility in decision conservativeness and exposes the system to greater downside risk.

  \subsection{Impact of Prediction Set Scaling Parameter}

We conduct a sensitivity analysis to examine how the conformal controller responds to variations in the prediction set scaling parameter $\sigma$ in Eq.~\eqref{eq:cc_set} under different levels of forecast accuracy. A larger $\sigma$ produces wider prediction sets and therefore more conservative dispatch decisions, whereas a smaller $\sigma$ yields narrower prediction sets and more aggressive actions driven by higher confidence in the forecasted opportunity value.

Under good forecast quality (Figure~\ref{fig:sigma}(\subref{fig:sigma_good})), the results are similar to that in Figure~\ref{fig:rho}(\subref{fig:rho_good}).

Under poor forecast quality (Figure~\ref{fig:sigma}(\subref{fig:sigma_bad})), an overly narrow prediction set (small $\sigma$) leads to significant increases in both cumulative risk $\hat R_T$ and realized profit loss, as the controller becomes overconfident and vulnerable to prediction errors, violating the conditions of Proposition~\ref{prop:optimal_width}. Conversely, given overly wide prediction sets (large $\sigma$), the cumulative risk $\hat R_T$ remains low, but the realized profit loss for CC-prediction-err increases. This effect is likely due to inappropriate estimate of heuristic conservativeness calibration function~$f$.

  \begin{figure}[t]
    \centering
    \begin{subfigure}[b]{0.49\textwidth}
        \centering
        \includegraphics[width=\textwidth]{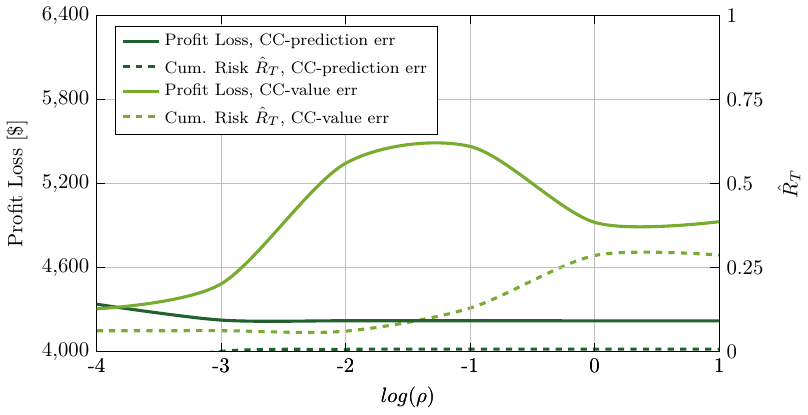}
        \caption{$R^2 = 0.4$.}
        \label{fig:rho_good}
    \end{subfigure}
  \hfill
    \begin{subfigure}[b]{0.49\textwidth}
        \centering
        \includegraphics[width=\textwidth]{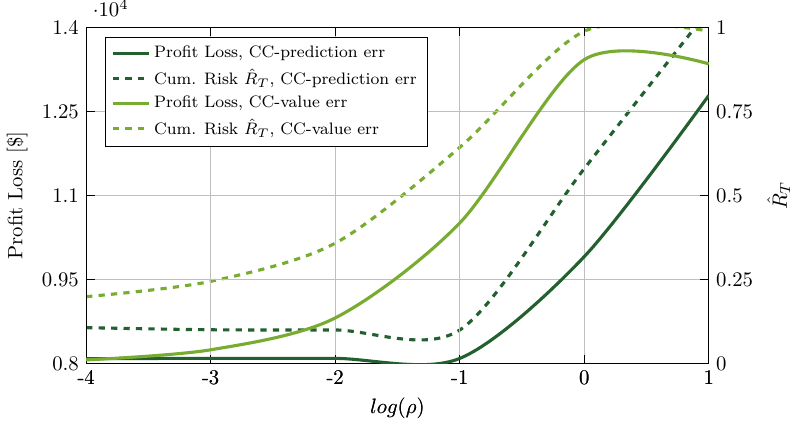}
        \caption{$R^2 = -0.4$.}
        \label{fig:rho_bad}
    \end{subfigure}
    \caption{Impact of learning rate $\rho$ considering good-accuracy ($R^2 = 0.4$) and poor-accuracy  ($R^2 = -0.4$) forecasters.}
    \label{fig:rho}
  \end{figure}

  \begin{figure}[t]
    \centering
    \begin{subfigure}[b]{0.49\textwidth}
        \centering
        \includegraphics[width=\textwidth]{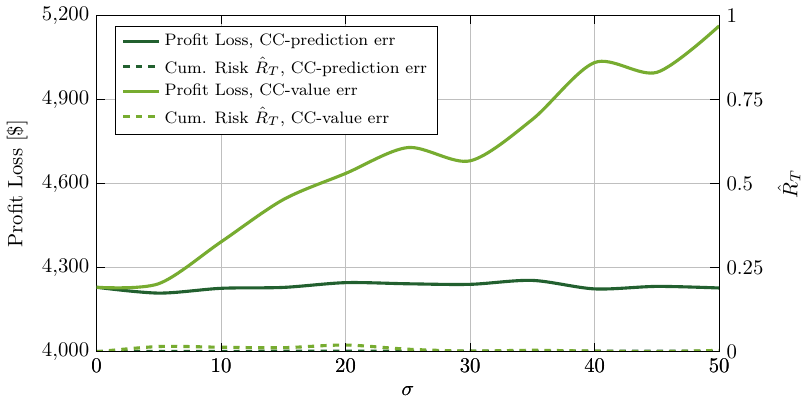}
        \caption{$R^2 = 0.4$.}
        \label{fig:sigma_good}
    \end{subfigure}
  \hfill
    \begin{subfigure}[b]{0.49\textwidth}
        \centering
        \includegraphics[width=\textwidth]{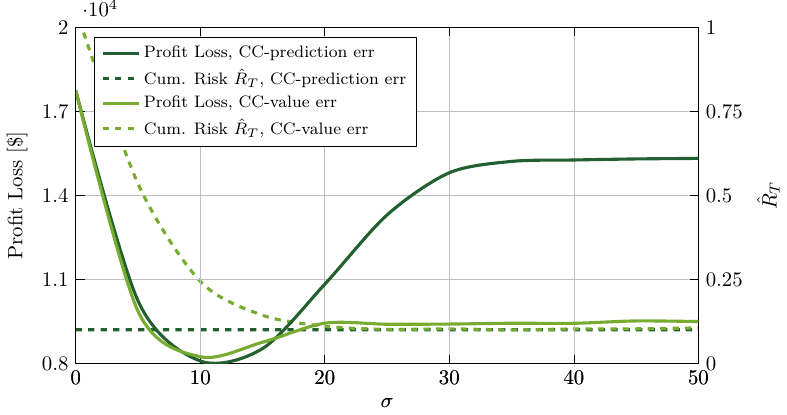}
        \caption{$R^2 = -0.4$.}
        \label{fig:sigma_bad}
    \end{subfigure}
    \caption{Impact of prediction set sensitivity parameter $\sigma$ considering good-accuracy ($R^2 = 0.4$) and poor-accuracy  ($R^2 = -0.4$) forecasters.}
    \label{fig:sigma}
  \end{figure}

}
\end{document}